\documentclass[12pt]{article}
\pdfoutput=1
\usepackage[utf8]{inputenc}
\usepackage[DIV=13]{typearea}
\usepackage{ragged2e}
\usepackage{amsmath,amsfonts,bbm,mathrsfs,amssymb}
\usepackage{bbold}
\usepackage{slashed,cancel}
\usepackage[usenames,dvipsnames]{xcolor}
\usepackage{cite}
\usepackage{bm} 
\usepackage{hyperref}
\hypersetup{colorlinks=true,urlcolor=Magenta,anchorcolor=blue,citecolor=blue,filecolor=blue,
            linkcolor=Magenta,menucolor=blue, linktocpage=true,pdfproducer=medialab}
\usepackage[english]{babel}
\usepackage{catchfile}
\usepackage{indentfirst}
\usepackage{graphicx}
\usepackage{float}
\usepackage{enumerate}
\usepackage{caption}
\usepackage{subfigure}
\usepackage{multirow}
\usepackage{tikz-feynman}
\usepackage[normalem]{ulem} 
\usepackage{booktabs}
\usepackage{physics}
\usepackage{nicefrac}
\usepackage[toc,page]{appendix}
\usepackage{pdflscape}

\newcommand{\email}[1]{\href{mailto:#1}{\tt #1}}

\numberwithin{equation}{section}

\newcommand{\blue}[1]{\color{blue} #1 \color{black}}

\newcommand{\magenta}[1]{\color{Magenta} #1 \color{black}}

\newcommand{\be}{\begin{equation}}
\newcommand{\ee}{\end{equation}}
\newcommand{\ba} {\begin{equation}\begin{aligned}}
\newcommand{\ea} {\end{aligned}\end{equation}}

\newcommand{\sL}{\mathscr{L}}

\newcommand{\cO}{\mathcal{O}}
\newcommand{\cM}{\mathcal{M}}

\newcommand{\cR}{\mathcal{R}}

\newcommand\subsetsim{\mathrel{%
  \ooalign{\raise0.2ex\hbox{$\subset$}\cr\hidewidth\raise-0.8ex\hbox{\scalebox{0.9}{$\sim$}}\hidewidth\cr}}}
\newcommand{\hc}{\text{h.c.}}
\newcommand{\ov}[1]{\overline{#1}}

\newcommand{\nn}{\nonumber}

\def\vs{{\textit vs.} }
\newcommand{\lrt}[1]{\left(#1\right)}
\newcommand{\lrq}[1]{\left[#1\right]}
\newcommand{\lrg}[1]{\left\{#1\right\}}

\newcommand{\eV}{\ \text{eV}}

\newcommand{\TeV}{\ \text{TeV}}
\newcommand{\GeV}{\ \text{GeV}}
\newcommand{\MeV}{\ \text{MeV}}

\def\tH{\widetilde{H}}

\begin{document} 
\renewcommand*{\thefootnote}{\fnsymbol{footnote}}
\begin{titlepage}

\vspace*{-1cm}
\flushleft{\magenta{IFT-UAM/CSIC-22-111}}
\\[1cm]

\begin{center}
\bf\LARGE \blue{
The Low-Scale Seesaw Solution to the}\\[2mm]
\boldmath
\bf\LARGE \blue{
$M_W$ and $(g-2)_\mu$ Anomalies}\\[4mm]
\unboldmath
\centering
\vskip .3cm
\end{center}
\vskip 0.5  cm
\begin{center}
{\large\bf Arturo de Giorgi}${}^{a)}$~\footnote{\email{arturo.degiorgi@uam.es}},
{\large\bf Luca Merlo}${}^{a)}$~\footnote{\email{luca.merlo@uam.es}}, and 
{\large\bf Stefan Pokorski}${}^{b)}$~\footnote{\email{Stefan.Pokorski@fuw.edu.pl}}
\vskip .7cm
{\footnotesize
${}^{a)}$ Departamento de F\'isica Te\'orica and Instituto de F\'isica Te\'orica UAM/CSIC,\\
Universidad Aut\'onoma de Madrid, Cantoblanco, 28049, Madrid, Spain
\vskip .2cm
${}^{b)}$ Institute of Theoretical Physics, Faculty of Physics,\\ 
University of Warsaw, Pasteura 5, PL 02-093, Warsaw, Poland
}
\end{center}
\vskip 2cm
\begin{abstract}
\justify
The recent CDF-II measurement of the $W$-boson mass shows a strong tension with the corresponding Standard Model prediction. Once active neutrino masses are explained in the context of the Low-Scale Seesaw mechanisms,  this tension can be resolved. We investigate the possibility of explaining the longstanding muon anomalous magnetic moment anomaly within the same frameworks.  We present a simplified extension of the Standard Model, accounting only for the second lepton generation, that describes a massive active neutrino and provides a combined solution to these anomalies.  The model is  renormalisable and introduces in the spectrum, beyond the sterile species of the Low-Scale Seesaw mechanism, only one pair of exotic vector-like leptons, doublets under the electroweak symmetry. We moreover discuss the extension of this model to the realistic three-family case.
\end{abstract}
\end{titlepage}
\setcounter{footnote}{0}

\pdfbookmark[1]{Table of Contents}{tableofcontents}
\tableofcontents

\renewcommand*{\thefootnote}{\arabic{footnote}}

\section{Introduction}
\label{sec:intro}

Among the scientific achievements in particle physics of the last century, the formulation of the Standard Model (SM) is one of the most relevant results. Its success culminated with the discovery of the Higgs boson in 2012 at the Large Hadron Collider~\cite{ATLAS:2012yve,CMS:2012qbp} and there has not been evidence of the existence of any other new particle till nowadays. 

On the other hand, the SM cannot be considered the ultimate theory of Nature.  It is lacking a mechanism  for the neutrino mass generation,   does not explain the baryon asymmetry of the Universe  and the  existence of Dark Matter,  and leaves aside the gravitational interactions. Moreover, several tensions are present between the SM predictions and the corresponding experimental determinations. One of the latest anomalies is associated with the mass of the $W$ gauge boson, $M_W$, that has been recently measured by the CDF II collaboration~\cite{CDF:2022hxs} with the best-achieved sensitivity,
\be
M_W=80.4335\pm0.0094\GeV\,,
\label{CFFIIMW}
\ee
showing a $7\sigma$ discrepancy relative to the SM prediction.\footnote{The CDF II measurement is by far the most precise one over the nine different determinations. As reported in Ref.~\cite{CDF:2022hxs}, neglecting the possible correlations, one could estimate the average among them obtaining $M_W=80.4242\pm0.0087\GeV$, that is only $1\sigma$ far from the CDF II result, as expected. For this reason, we will perform our analysis considering the value in Eq.~\eqref{CFFIIMW}, understanding that the conclusions would remain invariant using instead the average quantity.} 

Among the numerous proposals to explain such a tension, the class of the so-called Low-Scale Seesaw (SS) mechanisms~\cite{Kersten:2007vk,Abada:2007ux} represents a very appealing possibility~\cite{Blennow:2022yfm,Arias-Aragon:2022ats}, as it also provides a description for massive neutrinos. In the generic model of this type,  two kinds of exotic neutral leptons, $N_R$ and $S_R$ in the following, with opposite lepton numbers, are added to the SM spectrum. The number of the exotic neutral leptons of each type depends on the specific realisation: for example, in Ref.~\cite{Arias-Aragon:2022ats}, three $N_R$ and three $S_R$ have been considered in the exotic spectrum. In general, for an arbitrary number of $N_R$ and of $S_R$, it is convenient to adopt a compact notation for all the neutral leptons, SM and exotic, defining a multidimensional vector in the flavour space
\be
\chi\equiv(\nu_L,\, N_R^c,\, S_R^c)^T\,,
\label{LSSchiDef}
\ee
where $\nu_L$ stands for the neutral component of the EW lepton doublet $\ell_L$. The characteristic mass term for the Low-Scale SS setup reads
\be
-\sL_Y\supset\dfrac{1}{2}\ov{\chi}\cM_\chi\chi^c+\hc\,,
\label{GenericLowSSNeutralMassLag}
\ee
with
\be
\cM_\chi=
\begin{pmatrix}
0 & \dfrac{v}{\sqrt2} Y_N & \epsilon\dfrac{v}{\sqrt2} Y_S\\
\dfrac{v}{\sqrt2} Y_N^T & \mu' & \Lambda\\[4mm]
\epsilon\dfrac{v}{\sqrt2} Y^T_S & \Lambda^T & \mu \\
\end{pmatrix},
\label{genericLowSSmatrix}
\ee
where $v=246\GeV$ is the Higgs doublet $H$ vacuum expectation value (VEV), $Y_N$ and $Y_S$ are the Dirac Yukawa matrices that couple $\ell_L$ to $N_R$ and $S_R$ respectively, while $\Lambda$, $\mu$ and $\mu'$ are matrices in the flavour space of $N_R$ and $S_R$, and finally $\epsilon$ is a real parameter.

If lepton number conservation is taken as an exact symmetry, then the terms corresponding to $\mu$, $\mu'$ and $\epsilon$ are forbidden in the Lagrangian and then the neutrinos would remain massless. Allowing, however,  for an explicit soft violation  of the  lepton number conservation,  that is for non-zero values of the parameters $\mu$, $\mu'$ and $\epsilon$ (or some of them) and with $\Lambda$ dominating the other entries, the light neutrino mass matrix at tree-level reads
\be
m_\nu\simeq\dfrac{v^2}{2}\left[\left(Y_N\dfrac{1}{\Lambda^T}\mu\dfrac{1}{\Lambda}Y_N^T\right)-\epsilon\left(Y_S\dfrac{1}{\Lambda}Y_N^T+Y_N\dfrac{1}{\Lambda^T}Y^T_S\right)\right]\,.
\label{genericLowSSmnu}
\ee
Notice that $\mu'$ does not contribute to the neutrino masses at this expansion order.  Depending on which contribution dominates, a different name is used in the literature to refer to the specific SS mechanism: ``Inverse (ISS)''~\cite{Wyler:1982dd,Mohapatra:1986bd,Bernabeu:1987gr} if the first term in Eq.~\eqref{genericLowSSmnu}  is  mainly responsible for the neutrino masses; ``Linear (LSS)''~\cite{Malinsky:2005bi} otherwise.

In order to reproduce the atmospheric mass splitting $|\Delta m_\text{atm}^2|\sim2.5\times 10^{-3}\eV^{2}$~\cite{Esteban:2020cvm}, assuming $\Lambda\sim\cO(\text{TeV})$ and the Yukawa matrices with entries of $\cO(1)$, we can estimate the values for $\mu$ and $\epsilon$: $\mu\sim\text{KeV}$ in the ISS and $\epsilon\sim10^{-10}$ in the LSS. This is  the first interesting feature of these Low-Scale mechanisms: the masses of the light  neutrinos are explained through a soft breaking of the lepton number  conservation, with  small values of the  parameters $\epsilon$ and/or $\mu^{(\prime)}$, instead of  a  breaking by the large masses for the sterile species, such as in the traditional Type-I SS mechanism~\cite{Minkowski:1977sc,Gell-Mann:1979vob,Yanagida:1979as,Mohapatra:1979ia}. Indeed, the heavy neutral leptons can have masses at the TeV scale, which makes them possibly detectable at colliders. A second relevant aspect is that the Yukawa couplings are sizable and so  are the mixings between the sterile species and the active neutrinos. This is particularly interesting due to the significant induced deviation from the unitarity of the PMNS matrix~\cite{Langacker:1988ur,Antusch:2006vwa,Antusch:2014woa,Fernandez-Martinez:2016lgt}, which translates into a tree-level contribution to $M_W$ and allows to solve the CDF II anomaly, as discussed in Refs.~\cite{Blennow:2022yfm,Arias-Aragon:2022ats}. Indeed, any modification of the leptonic charged current implies an additional contribution to the muon $\beta$ decay, from which the value of the Fermi Constant $G_F$ (defined as the parameter that enters the Fermi Lagrangian) is computed: labelling with $G_\mu$ the parameter extracted from the muon lifetime,
\be
G_F=G_\mu(1+\Delta_{G}),
\ee
where $\Delta_{G}$ represents the generic deviation. Its effect appears in the prediction of $M_W$ as
\be
M_W=M_Z\sqrt{\dfrac12+\sqrt{\dfrac14-\dfrac{\pi\,\alpha\,(1-\Delta_G)}{\sqrt2\,G_\mu\,M_Z^2(1-\Delta r)}}}\,,
\ee
where $\alpha$ is the fine-structure constant, $M_Z$ is the mass of the $Z$ gauge boson and $\Delta r$ accounts for loop corrections. A value of $\Delta_G\sim5\times 10^{-3}$ would solve the tension in $M_W$.

Other observables are affected by the deviations in the leptonic charged current, such as the invisible $Z$ decay, various meson decays used to extract the values of the CKM matrix elements, and several decay width ratios that test the lepton flavour universality of the SM. As concluded in Ref.~\cite{Blennow:2022yfm}, the explanation of the CDF II anomaly is consistent with all these observables, but at the price of worsening the so-called Cabibbo Anomaly: the extracted value of the CKM entry $V_{ud}$ after the inclusion of the non-unitarity effects turns out to be larger than its actual value. We will ignore in this paper the Cabibbo anomaly, assuming that a different new physics may be responsible for its explanation.\footnote{Recently, an updated study by the UTfit collaboration has been published~\cite{UTfit:2022hsi}. The results seem to slightly reduce the significance of the Cabibbo anomaly.}

The main goal of this paper is to propose a possible modification of the Low-Scale SS models in order to solve the longstanding anomaly associated with the muon anomalous magnetic moment, $(g-2)_\mu$. It has been measured by the Muon $g-2$ collaboration at the Brookhaven National Laboratories~\cite{Muong-2:2006rrc} and more recently at Fermilab~\cite{Muong-2:2021ojo}, showing a combined $4.2\sigma$ tension with the SM result~\cite{Aoyama:2020ynm},
\be
\delta a_\mu
\equiv\dfrac{(g-2)_\mu^\text{exp}-(g-2)_\mu^\text{SM}}{2}
=\lrt{2.51\pm0.59}\times10^{-9}\,.
\label{DeltaamuExp}
\ee
It has to be mentioned that the BMW lattice collaboration~\cite{Borsanyi:2020mff} has recently presented a lattice result for the SM contributions to the hadronic vacuum polarization that would soften this tension. After that, other groups~\cite{Ce:2022kxy,Alexandrou:2022amy} seem to align with that result.  One may then wonder if QCD effects  not only explain the $(g-2)_\mu$ anomaly but also the one associated to the $M_W$ determination. However, as discussed in Ref.~\cite{Athron:2022qpo},  this is not the case as new physics is necessary to explain both of them simultaneously. Moreover, the lattice results are in tension with the $e^+e^-\to\text{hadrons}$ data~\cite{Crivellin:2020zul,Keshavarzi:2020bfy,Malaescu:2020zuc,Colangelo:2020lcg},  so the global situation associated to the $(g-2)_\mu$ remains unclear. While waiting for further calculations to establish a clear picture, we will adopt the result in Eq.~\eqref{DeltaamuExp} for the rest of the paper. 

It is well known that no  solution to the $(g-2)_\mu$ anomaly can be obtained with only sterile leptons. In several Refs.~\cite{Kannike:2011ng,Dermisek:2013gta,Arcadi:2021cwg,Lu:2021vcp,Guedes:2022cfy}, a broad analysis has been performed investigating which exotic fields beyond the SM (BSM) spectrum may play an interesting role in this respect. An approach that received attention in the last two years~\cite{Arkani-Hamed:2021xlp,Craig:2021ksw,DelleRose:2022ygn} consists of the introduction of vector-like leptons transforming as a doublet of the electroweak (EW) symmetry. Its attractiveness  resides in the absence of contributions to the $(g-2)_\mu$ suppressed by only two powers of the mass of the exotic states. The first relevant chirally enhanced term is suppressed by the fourth power of the mass. This represents a scenario where light new physics may be responsible for the $(g-2)_\mu$ anomaly. Notice that  this framework requires, beyond the exotic EW doublets, the presence of also leptonic sterile species in order to explain this anomaly. 

The question we want to answer in this paper is whether the introduction of additional vector-like lepton EW doublets, discussed in Refs.~\cite{Kannike:2011ng,Dermisek:2013gta,Arcadi:2021cwg,Lu:2021vcp,Guedes:2022cfy}, in the Low-Scale SS setups  may consistently explain the light active neutrino masses, the CDF II tension in $M_W$, and the $(g-2)_\mu$ anomaly. This construction would then represent a minimal setup where all the additional fields with respect to the SM spectrum are strictly necessary. With respect to the past literature, we revisit the analysis of the $(g-2)_\mu$, pointing out any difference in signs and factors and investigating new part of the parameter space, while explaining for the first time in this context the lightness of the active neutrinos and the $M_W$ CDF II measurement.

We will first proceed in Sect.~\ref{sec:OneGen} with the formulation of a minimal and simplified one-generation scenario that describes only the muon and the muonic neutrino. This will be realised by introducing only one $N_R$, one $S_R$ and one pair of vector-like EW doublets of exotic leptons. The model is a renormalisable extension of the Standard Model with all  the terms allowed by the Standard Model symmetries present in the Lagrangian. In Sect.~\ref{sec:PhenoSection}, we present the relevant phenomenology, distinguishing between observables that receive contributions at tree-level from those that do at loop-level. These two sections provide a proof of concept of the existence of a framework with massive neutrinos, where the CDF II $M_W$ tension and the $(g-2)_\mu$ anomaly can be simultaneously solved. In Sect.~\ref{sec:ThreeGen}, we comment on possible generalisations to account for the three generations of fermions, discussing the advantages and the disadvantages of each of them. Finally, we will conclude in Sect.~\ref{sec:Conc}.

\section{Formulation of the one-generation model}
\label{sec:OneGen}

This section is devoted to the description and discussion of the simplified model that only treats the second lepton generation. The lepton and scalar sector of the framework is defined in Tab.~\ref{Tab.Case1} together with its transformation properties under the gauge EW symmetry and its lepton charges. 

\begin{table}[h!]
\centering
\begin{tabular}{c|ccc|} 
&  $SU(2)_L $ & $U(1)_Y$ & $U(1)_L$\\ 
\hline 
\hline 
&&&\\[-3mm]
$\ell_L$  	& $\bf 2$ 		& $-1/2$    & 1 \\
$\mu_R$   	& 1  			& 1	        & 1	\\
$H$       	& $\bf 2$ 		& $+1/2$ 	& 0 \\
\hline
&&&\\[-3mm]
$N_R$  		& 1  		    & 1 		& 1  \\
$S_R$    	& 1  			& 1 		& -1 \\
\hline
&&&\\[-3mm]
$\psi_L$   	& $\bf2$  		& $-1/2$	& 1	\\
$\psi_R$    & $\bf2$ 		& $-1/2$  	& 1 \\
\hline
\end{tabular}
\caption{\em Transformation properties of the SM leptons $\ell_L$ and $\mu_R$, the Higgs doublet $H$, the sterile neutrinos $N_R$ and $S_R$, and of the leptonic vector-like EW doublet $\psi$ under the gauge EW symmetry and their lepton charges.}
\label{Tab.Case1}
\end{table}

The notation refers to the second lepton generation and therefore $\ell_L$ stands for the leptonic EW doublet containing the left-handed (LH) muon and muonic neutrino, while $\mu_R$ is the right-handed (RH) muon. $N_R$ and $S_R$ are two fermionic EW singlets  with opposite lepton charges, while $\psi_L$ and $\psi_R$ constitute a leptonic vector-like EW doublet. Finally, $H$ is the Higgs EW doublet. 

The corresponding mass Lagrangian can be written as 
\be
\begin{aligned}
-\sL_Y=&\phantom{+}\ov{\ell_L}HY_\mu \mu_R+
\ov{\ell_L}\tH Y_N N_R+
\epsilon\ov{\ell_L}\tH Y_S S_R+\dfrac12\mu'\ov{N_R^c}N_R+
\dfrac12\mu\ov{S_R^c}S_R+
\Lambda\ov{N_R^c}S_R+\\
&+Y_R\ov{\psi_L}H\mu_R+
Y_V\ov{S_R^c}\tH^\dag\psi_R+
Y'_V\ov{\psi_L}\tH N_R+
M_\psi\ov{\psi_L}\psi_R+
M_L\ov{\ell_L}\psi_R+
\hc\,,
\end{aligned}
\label{1GenLag}
\ee
where $\tH\equiv i\sigma_2H^\ast$, with $\sigma_2$ the second Pauli matrix, and all the terms respect the  lepton number conservation, except for those proportional to $Y_S$, $\mu$ and $\mu'$ which ensure the Low-Scale SS mechanism discussed in the Introduction. Without loss of generality and in order to keep the construction as minimal as possible, we will neglect the $\mu'$ term in what follows. It does not intervene in the determination of the neutrino masses at the order considered and, secondly, it is expected to be as small as $\mu$ and therefore negligible for the $(g-2)_\mu$ contributions, which is the main topic we want to address.\footnote{Our model is a renormalisable construction with a Lagrangian that includes all the terms invariant under the SM gauge symmetry. This differs from the setup considered in Ref.~\cite{Arkani-Hamed:2021xlp}, where the tree-level muon Yukawa term has been neglected: as a consequence, that framework is not renormalisable and a physical cut-off at about $10^5\GeV$ has been considered.}

Once the EW symmetry is spontaneously broken by the Higgs VEV, the mass terms can be compactly rewritten as
\be
-\sL_Y\supset
\dfrac{1}{2}\ov{\chi}\cM_\chi\chi^c+
\ov{\zeta_L}\cM_\zeta\zeta_R+
\hc\,,
\ee
where the neutral lepton multiplet $\chi$ and the charged one $\zeta$ are defined as 
\be
\chi\equiv(\nu_L,\, N_R^c,\, S_R^c,\,\psi^0_L,\,\psi_R^{0c})^T\,,\qquad
\zeta\equiv(\mu,\,\psi^-)^T\,,
\ee
generalising the definition of $\chi$ in Eq.~\eqref{LSSchiDef} to include the neutral components of $\psi$. The mass matrices $\cM_\chi$ and $\cM_\zeta$ are then written as
\begin{equation}
\cM_\chi= 
    \left(
    \begin{matrix}
        0   & m_N       & \epsilon\, m_S       & 0         & M_L \\
        m_N & 0         & \Lambda   & m_{V'}    & 0 \\
        \epsilon\, m_S & \Lambda   & \mu       & 0         & m_V \\
        0   & m_{V'}    & 0         & 0         & M_\psi  \\
        M_L & 0         & m_V       & M_\psi    & 0 \\
    \end{matrix}
    \right)\,,\qquad
\cM_\zeta= 
\left(
\begin{matrix}
m_\mu & M_L\\
m_R & M_\psi
\end{matrix}
\right)\,,
\end{equation}
where we use a shortcut notation for the product of the EW VEV and a Yukawa coupling, such that $m_i\equiv v Y_{i}/\sqrt2$.

Before entering  into the details of the (block) diagonalisation of these mass matrices, a few comments are in order. If the only non-vanishing entries would be those with $\Lambda$ and $M_\psi$, then $(N_R,\,S_R)$, $(\psi^0_L,\,\psi^0_R)$ and $(\psi^-_L,\,\psi^-_R)$ would be three massive Dirac pairs, while the neutrino and the muon would remain massless. The introduction of $M_L$ does not change this feature: the determinant of the two mass matrices would still be zero and therefore the lightest neutral and charged states would still be massless. However, the presence of $M_L$ induces a redefinition of the mass for the Dirac pairs $(\psi^0_L,\,\psi^0_R)$ and $(\psi^-_L,\,\psi^-_R)$, and more importantly leads to a mixing between the components of the LH fields $\ell_L$ and $\psi_L$. As we will see later, this results in the muon and the neutrino being composite states, whose level of compositeness depends on the hierarchy between $M_\psi$ and $M_L$. Once the terms proportional to the EW VEV and $\mu$ are considered, the muon and the neutrino acquire masses. With respect to the traditional Low-Scale SS mechanisms, the introduction of the EW doublet $\psi$, besides making the neutrino composite, does not modify either the expression for the neutrino mass or that for the mixing between the light active neutrino and the heavy species. We thus expect that the neutrino phenomenology for this model will remain essentially unmodified with respect to the one of the traditional Low-Scale SS mechanisms.

Given the interesting physical impact of the presence of $M_L$, it is illustrative to discuss an intermediate step in the mass matrix diagonalisation, that we will refer to with a tilde in the different quantities. This will be very useful for discussing the phenomenology of the model as the different observables can be written in a compact form in terms of the tilde parameters. 

The mass $M_L$ may be large and therefore the diagonalisation procedure would require a large rotation. The fields in the tilde basis, where $M_L$ does not appear in the off-diagonal entries of the mass matrices, read: for the charged leptons
\ba
\widetilde\mu_L& \equiv  \cos{\theta}\, \mu_L -\sin{\theta}\,\psi^-_L \qquad\qquad
    \widetilde\mu_R\equiv \mu_R\\ 
\widetilde\psi^-_L& \equiv  \sin{\theta}\,\mu_L +\cos{\theta}\,\psi^-_L\qquad\qquad
\widetilde\psi^-_R\equiv \psi^-_R
\ea
and for the neutral leptons
\ba
\widetilde\nu_L&\equiv \cos{\theta}\,\nu_L-\sin{\theta}\,\psi^0_L\\
\widetilde{N}_R&\equiv N_R\qquad\qquad
&&\widetilde{S}_R\equiv S_R\\
\widetilde\psi_L^0&\equiv 
\sin{\theta}\,\nu_L+
\cos{\theta}\,\psi^0_L\qquad\qquad
&&\widetilde\psi_R^{0}\equiv \psi^0_R\,.
\ea
As we can see, only the LH components of the two EW doublets are affected by this redefinition while the other fields remain the same. In these expressions, $\theta$ is a mixing angle defined as
\be
\cos{\theta}\equiv \dfrac{M_\psi}{\widetilde{M}_\psi}\,, \qquad\qquad \sin{\theta}\equiv \dfrac{M_L}{\widetilde{M}_\psi}\,,\qquad\qquad
\text{with}\qquad 
\widetilde{M}_\psi \equiv \sqrt{M_\psi^2+M_L^2}\,.
\label{ThetaDefinitionToy}
\ee
In the case in which $M_L$ and $M_\psi$ are taken of the same order of magnitude then the angle is close to $45^\circ$, while once $M_L$ is negligible (dominant) with respect to $M_\psi$ then $\sin\theta\approx0$ ($\cos\theta\approx0$). We will further discuss these three cases later in this section.  

Finalising the diagonalisation, all the fields are redefined and the light active neutrino gets mass.  Denoting by a ``hat'' the quantities in the mass basis, the charged lepton masses are given by
\be
\widehat{m}_\mu = \widetilde{m}_\mu\left[1-\dfrac{1}{2}\left(\dfrac{\widetilde{m}_R}{\widetilde{M}_\psi}\right)^2\right]\,,\qquad\qquad
\widehat{m}_{\psi^-} =\widetilde{M}_\psi \left[1+\dfrac{1}{2}\left(\dfrac{\widetilde{m}_R}{\widetilde{M}_\psi}\right)^2\right] \,,
\ee
where we used as definitions 
\be
\widetilde{m}_\mu = m_\mu \cos{\theta} - m_{R}\sin{\theta}\,,\qquad\qquad
\widetilde{m}_R = m_{R} \cos{\theta} + m_{\mu}\sin{\theta}\,,
\ee
and we neglected terms that have a relative suppression at least equal to $(v/\widetilde{M}_\psi)^2$ with respect to the expressions of the masses. The corresponding mass eigenstates are defined by
\be
\widehat{\zeta}:\qquad
\begin{cases}
\widehat\mu_L =  \widetilde{\mu}_L \,,\qquad
&\widehat\psi^-_L = \widetilde{\psi}^-_L \,,\\[2mm]
\widehat\mu_R = \widetilde{\mu}_R - \left(\dfrac{\widetilde{m}_R}{\widetilde{M}_\psi}\right) \widetilde{\psi}^-_R \,,\qquad 
&\widehat\psi^-_R =  \widetilde{\psi}^-_R + \left(\dfrac{\widetilde{m}_R}{\widetilde{M}_\psi}\right) \widetilde{\mu}_R\,,
\end{cases}
\label{ZetaMassEig}
\ee
where now the neglected terms are more suppressed by only a $v/\widetilde{M}_\psi$ power.

Analogously, for the neutral lepton sector, the final expressions for the masses are
\ba
\widehat{m}_\nu &=\dfrac{\mu\,\widetilde{m}_N^2}{\Lambda^2}-\dfrac{2\,\epsilon\,\widetilde{m}_N\, m_S \cos{\theta}}{\Lambda} \,,\\
\widehat{m}_{N_R} &= \Lambda+\dfrac{\mu}{2} + \dfrac{\widetilde{m}_N^2}{2 \Lambda}+ \dfrac{1}{4}\left[\dfrac{(m_V+\widetilde{m}_{V'})^2}{\Lambda-\widetilde{M}_\psi}+\dfrac{(m_V-\widetilde{m}_{V'})^2}{\Lambda+\widetilde{M}_\psi}\right]\,,\\
\widehat{m}_{S_R}&= \Lambda -\dfrac{\mu}{2}+ \dfrac{\widetilde{m}_N^2}{2 \Lambda}+ \dfrac{1}{4}\left[\dfrac{(m_V+\widetilde{m}_{V'})^2}{\Lambda-\widetilde{M}_\psi}+\dfrac{(m_V-\widetilde{m}_{V'})^2}{\Lambda+\widetilde{M}_\psi}\right]\,,\\
\widehat{m}_{\psi^0} &= \widetilde{M}_\psi -\dfrac{1}{4}\left[\dfrac{(m_V+\widetilde{m}_{V'})^2}{\Lambda-\widetilde{M}_\psi}-\dfrac{(m_V-\widetilde{m}_{V'})^2}{\Lambda+\widetilde{M}_\psi}\right]\,,
\label{FinalMassesNeutral}
\ea
while those for the mass eigenstates read
\be
\widehat{\chi}:\qquad
\begin{cases}
\widehat\nu_L = \widetilde\nu_L-\dfrac{\widetilde{m}_N}{\Lambda}\,\widetilde{S}^c_R\,,\\[2mm]
\widehat N_R = \dfrac{\widetilde{N}_R+\widetilde{S}_R}{\sqrt2} +\dfrac{\widetilde{m}_N}{\sqrt{2}\Lambda}\widetilde\nu_L^c 
+\dfrac{1}{2}\left[\dfrac{m_V+\widetilde{m}_{V'}}{\Lambda-\widetilde{M}_\psi} \dfrac{\widetilde{\psi}^{0c}_L+\widetilde{\psi}^0_R}{\sqrt2}-
\dfrac{m_V-\widetilde{m}_{V'}}{\Lambda+\widetilde{M}_\psi}\dfrac{\widetilde{\psi}^{0c}_L-\widetilde{\psi}^0_R}{\sqrt2}\right] \,,\\[2mm]
\widehat S_R = i\lrg{-\dfrac{\widetilde{N}_R-\widetilde{S}_R}{\sqrt2}+\dfrac{\widetilde{m}_N}{\sqrt{2}\Lambda}\widetilde\nu_L^c-\dfrac{1}{2}\left[\dfrac{m_V-\widetilde{m}_{V'}}{\Lambda+\widetilde{M}_\psi} \dfrac{\widetilde{\psi}^{0c}_L+\widetilde{\psi}^0_R}{\sqrt2}-
\dfrac{m_V+\widetilde{m}_{V'}}{\Lambda-\widetilde{M}_\psi}
\dfrac{\widetilde{\psi}^{0c}_L-\widetilde{\psi}^0_R}{\sqrt2}\right]}  \,,\\[2mm]
\widehat\psi^0_L = \dfrac{\widetilde{\psi}^0_L+\widetilde{\psi}^{0c}_R}{\sqrt2} -
\dfrac{1}{2} \left[\dfrac{m_V+\widetilde{m}_{V'}}{\Lambda-\widetilde{M}_\psi} \dfrac{\widetilde{N}^c_R+\widetilde{S}^c_R}{\sqrt2}+\dfrac{m_V-\widetilde{m}_{V'}}{\Lambda+\widetilde{M}_\psi}\dfrac{\widetilde{N}^c_R-\widetilde{S}^c_R}{\sqrt2}\right]\,,\\[2mm]
\widehat\psi^0_R = i\lrg{-\dfrac{\widetilde{\psi}^{0c}_L-\widetilde{\psi}^0_R}{\sqrt2}-\dfrac{1}{2} \left[\dfrac{m_V-\widetilde{m}_{V'}}{\Lambda+\widetilde{M}_\psi} \dfrac{\widetilde{N}_R+\widetilde{S}_R}{\sqrt2}+\dfrac{m_V+\widetilde{m}_{V'}}{\Lambda-\widetilde{M}_\psi}\dfrac{\widetilde{N}_R-\widetilde{S}_R}{\sqrt2}\right]}\,,
\end{cases}
\label{ChiMassEig}
\ee
where we defined
\be
\widetilde{m}_N = m_N \cos{\theta} - m_{V'}\sin{\theta}\,,\qquad\qquad
\widetilde{m}_{V'} = m_{V'} \cos{\theta} + m_{N}\sin{\theta}\,
\label{mNandmVpDefs}
\ee
and we neglected higher order terms: some of them follow the same pattern as the charged states, that is $\lrt{v/(\Lambda,\,\widetilde{M}_\psi)}^2$ for the masses and $v/(\Lambda,\,\widetilde{M}_\psi)$ for the eigenstates; while others are proportional to $(\mu,m_S)/(\Lambda,\widetilde{M}_\psi)$. This holds for all the expressions, except for $\widehat{m}_\nu$, where the terms with $\mu$ and $m_S$ are the leading ones. 
Requiring that the neglected terms are smaller than the $20\%$ of the shown contributions to the masses translates into a lower bound for $\Lambda$ and $\widetilde M_\psi$, such that
\be
\Lambda,\,\widetilde M_\psi\gtrsim500\GeV\,.
\label{BoundHNLMassesTheo}
\ee
This consistency constraint will be imposed in the following analysis.
Moreover, the ``$i$'' in front of the third and fifth lines guarantees that the corresponding expressions for the masses are defined as positive. Special care is required for the active neutrinos: in case the LSS contribution proportional to $\epsilon$ dominates, then the mass would turn negative and then we should redefine $\widehat\nu_L$ introducing the dependence of an ``$i$'' factor. Finally, to match with the literature of the Low-Scale Seesaw, the mixing between active neutrinos and the heavy neutral leptons responsible for their masses, defined as $\widehat\nu_L\simeq\widetilde\nu_L-\Theta\,\widetilde{S}^c_R$, is given by
\be
\Theta=\dfrac{\widetilde{m}_N}{\Lambda}
\ee
and it will be useful for discussing the direct searches of the heavy states at colliders.

Moreover, it is interesting to consider three different scenarios depending on the hierarchy between $M_L$ and $M_\psi$:
\begin{description}
\item[$\mathbf{M_L\ll M_\psi}$:] in this case, $\sin\theta\sim0$ and the main contribution to the muon mass is due to the muon Yukawa term $Y_\mu$ and the muon field almost coincides with the elementary field $\mu$, such as in the SM.
\item[$\mathbf{M_L\gg M_\psi}$:] this is the opposite case with respect the previous one, where $\cos\theta\sim0$ and then the main contribution to the muon mass is now the $Y_R$ Yukawa term. Correspondingly, the LH component of the muon field coincides with the exotic field $\psi^-_L$, while the RH component can be almost identified with the SM $\mu_R$.
\item[$\mathbf{M_L\sim M_\psi}$:] in this intermediate case, the muon mass receives sizable contributions from both $Y_\mu$ and $Y_R$ Yukawa terms and the LH component of the muon field is a composite state of $\mu_L$ and $\psi^-_L$.
\end{description}

\section{Phenomenology of the one-generation model}
\label{sec:PhenoSection}

The presence of the exotic leptons can be tested both with deviations from the SM predictions and with direct searches at colliders. The analysis on indirect signals requires the effective Lagrangian describing the SM lepton couplings with the SM gauge bosons and with one physical Higgs $h$ that reads
\begin{align}
\sL_\text{SM} \supset & -\dfrac{h}{v}\left[ \widetilde{m}_\mu\left(1-\dfrac{3}{2}\dfrac{\widetilde{m}_R^2}{\widetilde{M}_\psi^2}\right)\,\ov{\widehat\mu_L}\widehat\mu_R+
\left(\dfrac{\mu\,\widetilde{m}_N^2}{\Lambda^2}-\dfrac{2\,\epsilon\,\widetilde{m}_N\, m_S \cos{\theta}}{\Lambda}\right)\,\ov{\widehat\nu_L}\widehat\nu_L^c+ \hc\right]+\nn\\
&+e\, \ov{\widehat\mu}\,\slashed{A}\, \widehat\mu -\dfrac{g_L}{\sqrt{2}}\left[\left(1-\dfrac{\widetilde{m}_N^2}{2\Lambda^2}\right)\, \ov{\widehat\nu_L}\,\slashed{W}^+\,\widehat\mu_L+ \hc\right]+
\label{eq:Lagrangian-SM}
\\
&-\dfrac{\sqrt{g_L^2+g_Y^2}}{2}\left[\left(1-\dfrac{\widetilde{m}_N^2}{\Lambda^2}\right)\,\ov{\widehat\nu_L}\,\slashed{Z}\, \widehat\nu_L-
\cos{2\theta_W}\widehat{\mu}\,\slashed{Z}\, \widehat{\mu}+\left(1-\dfrac{\widetilde{m}_R^2}{\widetilde{M}_\psi^2}\right)\widehat{\mu}_R\,\slashed{Z}\, \widehat{\mu}_R
\right]\nn\,,
\end{align}
where $g_{L(Y)}$ is the $SU(2)_L(U(Y))$ gauge-coupling and we neglected the Higgs coupling to the active neutrino as it does not lead to any relevant phenomenology.

It is useful to perform the matching with the phenomenological Lagrangian describing the same interactions,
\begin{align}
\sL_\text{SM}^\text{eff.}\supset
&-\dfrac{h}{v}
\lrq{
m^\text{exp}_\mu\, \kappa_\mu\, \ov{\widehat\mu}\, \widehat\mu + i\,m^\text{exp}_\mu\, \tilde{\kappa}_{\mu}\, \ov{\widehat\mu}\, \gamma_5\, \widehat\mu
} -\dfrac{g_L}{\sqrt{2}}\lrq{(1+\delta g^{W\mu}_L)\,\ov{\widehat\nu_L}\, \slashed{W}^+\, \widehat\mu_L+\hc}+
\label{eq:Lagrangian-effective}
\\
&-\sqrt{g_L^2+g_Y^2}\lrq{\sum\limits_{f=\mu,\nu}\lrq{\lrt{T^3_f-s_{\theta_W}^2 Q_f}+\delta g^{Zf}_L}\ov{\widehat{f}_L}\,\slashed{Z}\,f_L+ \lrt{s_{\theta_W}^2+\delta g^{Z\mu}_R}\ov{\widehat{\mu}_R}\,\slashed{Z}\,\widehat{\mu}_R}\,,
\nn
\end{align}
where $m^\text{exp}_\mu$ is the experimental value of the muon mass, $T^3_\nu=+1/2$ and $T^3_\mu=-1/2$, while $\delta g$ are the deviations from the corresponding SM values, and $\kappa_\mu$ and $\tilde{\kappa}_\mu$ represent the deviations from the SM (real and imaginary) values of the muon Yukawa. 

Comparing Eqs.~\eqref{eq:Lagrangian-SM} and \eqref{eq:Lagrangian-effective}, we find the following expression of the effective quantities in terms of the parameters of the model:
\be
\begin{gathered}
\kappa_\mu=1-\dfrac{3}{2}\dfrac{\widetilde{m}_R^2}{\widetilde{M}_\psi^2}\,,
\hspace{4cm}
\tilde\kappa_\mu=0\,,\\
\delta g^{W\mu}_L=\delta g^{Z\nu}_L=-\dfrac{\widetilde m_N^2}{2\Lambda^2} \,,
\qquad
\delta g^{Z\mu}_L=0\,,
\qquad
\delta g^{Z\mu}_R=-\dfrac{\widetilde{m}_R^2}{2\widetilde{M}_\psi^2}\,.
\end{gathered}
\label{MatchingWithPhenoLag}
\ee

On the other hand, in order to discuss direct searches at colliders, we need the Lagrangian describing the interactions of the physical Higgs and the SM gauge bosons with one light lepton and one heavy:
\begin{align}
\sL_h \supset & -\dfrac{h}{v}
\left\{\widetilde{m}_R\left(1-\dfrac{\widetilde{m}_R^2-2\widetilde{m}_\mu^2}{2\widetilde{M}_\psi^2}\right)\ov{\widehat\mu_R}\widehat\psi^-_L+
\dfrac{\widetilde{m}_\mu\widetilde{m}_R}{\widetilde{M}_\psi}\ov{\widehat\mu_L}\,\widehat\psi^-_R+\right. 
\label{hLagMix}\\
&\hspace{1cm}+\dfrac{\widetilde{m}_N}{2}\,\left[1-\dfrac{\widetilde{m}_N^2}{\Lambda^2}+\dfrac{m_V \widetilde{m}_{V'}}{\Lambda\widetilde{M}_\psi} -\dfrac{m_V}{\Lambda}\left(\dfrac{m_V \Lambda + \widetilde{m}_{V'}\widetilde{M}_\psi}{\Lambda^2-\widetilde{M}_\psi^2}\right)+\right.\nn\\
&\hspace{2.5cm}\left.-\dfrac{1}{2}\left(\dfrac{m_V \Lambda + \widetilde{m}_{V'}\widetilde{M}_\psi}{\Lambda^2-\widetilde{M}_\psi^2}\right)^2\right]\left(\ov{\widehat\nu_L^c}\dfrac{\widehat{N}_R^c-i\widehat{S}_R^c}{\sqrt{2}}+\ov{\widehat\nu_L}\dfrac{\widehat{N}_R+i\widehat{S}_R}{\sqrt{2}}\right)+\nn\\
&\hspace{1cm}\left.-\dfrac{\widetilde{m}_N}{2}\,\left[\dfrac{m_V}{\Lambda}+\dfrac{m_V \Lambda + \widetilde{m}_{V'}\widetilde{M}_\psi}{\Lambda^2-\widetilde{M}_\psi^2}\right]\left(\ov{\widehat\nu_L^c}\dfrac{\widehat{\psi}_L^0+i\widehat{\psi}^{0c}_R}{\sqrt{2}}+\ov{\widehat\nu_L}\dfrac{\widehat{\psi}_L^{0c}-i\widehat{\psi}^{0}_R}{\sqrt{2}}\right)\right\}+\hc\nn\\
\sL_Z \supset & -\dfrac{\sqrt{g_L^2+g_Y^2}}{2}Z_\mu\left\{\dfrac{\widetilde{m}_R}{\widetilde{M}_\psi}\, \ov{\widehat\mu_R}\gamma^\mu\widehat\psi^-_R+\dfrac{\widetilde{m}_N}{\Lambda}\,\ov{\widehat{\nu}_L}\gamma^\mu\dfrac{\widehat{S}_R^c-i \widehat{N}_R^c}{\sqrt{2}}+ \right.
\label{ZLagMix}\\
&\hspace{2cm}\left.+\dfrac{1}{2}\left[\dfrac{m_V+\widetilde{m}_{V'}}{\Lambda-\widetilde{M}_\psi}-\dfrac{m_V-\widetilde{m}_{V'}}{\Lambda+\widetilde{M}_\psi}\right]\ov{\widehat{\nu}_L}\gamma^\mu\dfrac{i \widehat{\psi}^0_L+\widehat{\psi^{0c}_R}}{\sqrt{2}}\right\}\nn
\end{align}
\begin{align}
\sL_W \supset & -\dfrac{g_L}{\sqrt{2}}W^-_\mu
\left\{\dfrac{\widetilde{m}_N}{\Lambda}\,\ov{\widehat\mu_L}\gamma^\mu\dfrac{\widehat{N}_R^c+i\widehat{S}_R^c}{\sqrt{2}}+\right.
\label{WLagMix}\\
&\hspace{2.5cm}\left.-\dfrac{\widetilde{m}_R}{2\widetilde{M}_\psi}\left[\dfrac{m_V+\widetilde{m}_{V'}}{\Lambda-\widetilde{M}_\psi}+\dfrac{m_V-\widetilde{m}_{V'}}{\Lambda+\widetilde{M}_\psi}\right]\ov{\widehat\mu_R}\gamma^\mu\dfrac{\widehat{N}_R+i\widehat{S}_R}{\sqrt{2}}+\right.\nn\\
&\hspace{2.5cm}\left.-\left[\dfrac{\widetilde{m}_\mu \widetilde{m}_R}{\widetilde{M}_\psi^2}+\dfrac{\widetilde{m}_N}{2\widetilde{M}_\psi}\left(\dfrac{m_V+\widetilde{m}_{V'}}{\Lambda-\widetilde{M}_\psi}+\dfrac{m_V-\widetilde{m}_{V'}}{\Lambda+\widetilde{M}_\psi}\right)\right]\ov{\widehat\mu_L}\gamma^\mu\dfrac{\widehat{\psi}_L^0-i\widehat{\psi}_R^{0c}}{\sqrt{2}}+\right.\nn\\
&\hspace{2.5cm}\left.-\dfrac{\widetilde{m}_R}{\widetilde{M}_\psi}\,\ov{\widehat\mu_R}\gamma^\mu\dfrac{\widehat{\psi}_L^{0c}-i\widehat{\psi}_R^{0}}{\sqrt{2}}\right\}+\hc\,.\nn
\end{align}

We can now proceed with the phenomenological analysis, dividing the discussion between observables receiving contributions at tree level and those at $1$-loop level.

\subsection{Relevant phenomenology at tree-level}
\label{sec:PhenoTreeLevel}

\subsubsection*{Colliders bounds on Higgs couplings}
The precise measurement of the Higgs couplings to fermions became a primordial goal after the Higgs discovery and ATLAS and CMS experiments have reported numerous results. The most recent combinations of different Higgs signal strengths have been recently released in Refs.~\cite{ATLAS:2019nkf} (ATLAS) and \cite{CMS:2018uag} CMS, using data at $\sqrt s=13\TeV$. In addition, both collaborations reported in Refs.~\cite{ATLAS:2020fzp} and \cite{CMS:2020xwi} the observations of Higgs decays into a pair of opposite-sign muons, in collisions at $\sqrt{s}=13\TeV$. Using these collider data, Ref.~\cite{Alonso-Gonzalez:2021tpo,Bahl:2022yrs,Brod:2022bww} performed a global fit obtaining a bound on a combination of $\kappa_\mu$ and $\tilde\kappa_\mu$, 
\be
0.36 \lesssim \kappa^2_\mu+\tilde\kappa^2_\mu\lesssim 1.85\,,
\ee
that, given the matching in Eq.~\eqref{MatchingWithPhenoLag}, translates into a bound on the ratio $\widetilde{m}^2_R/\widetilde{M}^2_\psi$,
\be
0.6 \lesssim \kappa_\mu\lesssim 1.36\quad\Longrightarrow\quad
\dfrac{\widetilde{m}^2_R}{\widetilde{M}^2_\psi}\lesssim0.27\,.
\ee

\subsubsection*{\boldmath EW global fit bounds on $Z$ couplings}

A stronger constraint on this parameter ratio can be extracted from the bounds on the deviation of the $Z-\mu$ coupling from its SM prediction: the results of the EW global fit performed in Ref.~\cite{Breso-Pla:2021qoe} gives
\begin{equation}
        \delta g^{Z\mu}_L = (0.1\pm 1.2)\cross 10^{-3}\,, \qquad\qquad
        \delta g^{Z\mu}_R = (0.0\pm 1.4)\cross 10^{-3} \,,
\end{equation}
that translates at the $2\sigma$ level to
\begin{equation}
\dfrac{\widetilde{m}_R^2}{\widetilde{M}_\psi^2}<5.6\cross 10^{-3} \,.
\label{eq:exp_bounds}
\end{equation}

\subsubsection*{\boldmath CDF II $M_W$ Tension}
The same EW global fit also gives a bound on $\delta g^{W\mu}_L$, but the input data used do not take into consideration the recent CDF II measurement of the $W$ mass and for this reason, we will proceed with a dedicated discussion in what follows. The modification of the $W$ coupling has an impact on the computation of the  muon $\beta$ decay: assuming that the $W$ coupling with the first generation leptons is as in the SM, the decay rate of $\mu\to e\ov{\nu}\nu$ reads
\begin{equation}
    \Gamma_\mu \simeq \dfrac{m^{\text{exp}\,5}_\mu G_F^2}{192\,\pi^3}\lrt{1-\dfrac{\widetilde m_N^2}{2\Lambda^2}}^2\equiv\dfrac{m^{\text{exp}\,5}_\mu G_\mu^2}{192\,\pi^3}\,,
\end{equation}
where we recall that $G_F$ is the Fermi constant parameter as defined in the Fermi Lagrangian, $G_\mu$ is the corresponding experimental determination extracted from the muon lifetime (after correcting for $O(\alpha_{EM})$ radiative effects).  The relation between these two quantities is such that
\begin{equation}
    G_F 
    \simeq G_\mu \left(1+\dfrac{\widetilde m_N^2}{2\Lambda^2}\right) \,,
\end{equation}
and this implies a modification of the relation between the $W$ boson mass and the experimental determination of $G_\mu$,
\begin{equation}
    M_W\simeq M_Z\,
    \sqrt{
    \dfrac12+
    \sqrt{
    \dfrac14-
    \dfrac{\pi\,\alpha_\text{em}}{\sqrt2\,G_\mu\,M_Z^2\,\lrt{1-\Delta r}}
    \lrt{1-\dfrac{\widetilde m_N^2}{2\Lambda^2}}
    }
    }\,,
    \label{MWPrediction1Gen}
\end{equation}
in the on-shell scheme, where the tree-level formula for the sine of the Weinberg angle is promoted to the definition of the renormalised quantity:
\be
\sin^2\theta_W\equiv1-\dfrac{M_W^2}{M_Z^2}\,.
\ee
Considering the following numerical values of the input parameters~\cite{Workman:2022ynf}
\be
\begin{aligned}
m^\text{exp}_\mu&=105.6583755(23)\MeV\\
\alpha_\text{em}&=7.2973525693(11)\times10^{-3}\\
G_\mu&=1.1663787(6)\times10^{-5}\GeV^{-2}\\
M_Z&=91.1876(21)\GeV\\
\Delta r&=0.03657(21)(7)\,,
\end{aligned}
\ee
we can extract the range of values necessary to explain the recent CDF II measurement in Eq.~\eqref{CFFIIMW}: at the $2\sigma$ level, 
\be
\dfrac{\widetilde m_N^2}{\Lambda^2}\in [6.6,11.8]\cross 10^{-3} \,.
\label{MWRequiredToy}
\ee
This bound is consistent with the results shown in Refs.~\cite{Blennow:2022yfm,Arias-Aragon:2022ats}. Additional modifications to the $W$ mass appear at the loop level, but they are completely negligible in the considered parameter space and therefore we will neglect them. 

\subsubsection*{\boldmath Effective $N_\nu$ and LFU ratios}

A second bound on this combination of parameters can be obtained from the modification of the $Z$-decay into neutrinos.   It is constrained by the experimental determination of the invisible $Z$ decay rate. The analytic expression for the $Z$ decay rate into neutrinos, assuming that only the coupling with the muon neutrino is modified according to Eq.~\eqref{eq:Lagrangian-SM}, while those with the electron and tau neutrinos remain as in the SM, reads
\be
\Gamma_\text{$Z$-inv}\simeq\dfrac{G_\mu\,M_Z^3}{12\,\sqrt2\,\pi}\lrt{3-\dfrac{\widetilde m_N^2}{2\Lambda^2}}\equiv\dfrac{G_\mu\,M_Z^3\,N_\nu}{12\,\sqrt2\,\pi}\,,
\label{InvisibleZDecayToy}
\ee
where $N_\nu$ is the number of effective active neutrinos. The experimental determination of the latter is $N_\nu^\text{exp}=2.9963(74)$~\cite{Janot:2019oyi} and it provides the following bound on the combination of parameters of the model:
\be
\dfrac{\widetilde m_N^2}{\Lambda^2}<3.7\times10^{-2}\,,
\ee
compatible with the range of values in Eq.~\eqref{MWRequiredToy} required to explain the new $M_W$ measurement. 

Further constraints follow from pion,  kaon and  also tau decays: the relative  branching ratios of the decay of those particles to different lepton flavours are clean observables that test the lepton flavour universality,
\be
\cR^P_{\mu/e}\equiv\dfrac{\Gamma\lrt{P\to \mu\,\ov{\nu_\mu}}}{\Gamma\lrt{P\to e\,\ov{\nu_e}}}\,,\qquad\qquad
\cR^\tau_{\mu/e}\equiv\dfrac{\Gamma\lrt{\tau\to \mu\,\ov{\nu_\mu}\,\nu_\tau}}{\Gamma\lrt{\tau\to e\,\ov{\nu_e}\,\nu_\tau}}\,.
\label{LFVratiosDefinitions}
\ee
In all cases, the deviation due to New Physics (NP) reduces to the same combination of parameters as only the muon sector is affected: comparing with the experimental determinations~\cite{Bryman:2021teu},
\be
1-\dfrac{\widetilde m_N^2}{2\Lambda^2}\Big|_\pi=1.0010(9)\,,
\qquad
1-\dfrac{\widetilde m_N^2}{2\Lambda^2}\Big|_K=0.9978(18)\,,
\qquad
1-\dfrac{\widetilde m_N^2}{2\Lambda^2}\Big|_\tau=1.0018(14)\,.
\label{MesonRatiosLFUToy}
\ee
The strongest bound on the combination of the model parameter at the $2\sigma$ level is
\be
\dfrac{\widetilde m_N^2}{\Lambda^2}<1.6\times 10^{-3}\,,
\ee
slightly in tension with the preferred region to explain the CDF II anomaly. We expect that this tension can be resolved once extending this analysis to the three flavour case,  as will be discussed in the next sections.

\subsubsection*{Direct searches of heavy leptons}

The only couplings between heavy and SM fermions which are not suppressed are the ones involving the Higgs and are proportional to $\widetilde{m}_{N,R}$, as can be seen in Eq.~\eqref{hLagMix}. One would therefore expect sizeable contributions to off-shell Higgs-mediated processes.  The bounds on such couplings are extremely interesting,  but they go beyond the scope of this work and are left for more detailed and dedicated future analysis.

Concerning the couplings with $Z$ and $W$ gauge bosons, we can distinguish two main scenarios: i) $\Lambda<\widetilde M_\psi$, such that among the heavy states, the lightest are $\widehat{N}_R$ and $\widehat{S}_R$ (see Eq.~\eqref{FinalMassesNeutral}); ii) $\Lambda>\widetilde M_\psi$ and the lightest states are $\widehat{\psi}^{0,-}$. In the case i), the direct search strategy falls in the category of the standard Heavy Neutral Lepton (HNL) scenario (see Ref.~\cite{Abdullahi:2022jlv} for a recent summary). The present experimental bounds from CMS apply only for $\widehat{m}_{N_R,S_R}<1.2\TeV$ and they apply to $\widetilde m_N^2/\Lambda^2$, that is the combination of parameters that control the mixing between the light and heavy species both in the neutral and in the charged gauge currents: considering the smallest masses that $\widehat N_R$ and $\widehat S_R$ can take consistently with Eq.~\eqref{BoundHNLMassesTheo}, that is $\widehat{m}_{N_R,S_R}\simeq500\GeV$, the corresponding bound is weaker than the indirect searches listed above and reads
\be
\lrq{\dfrac{\widetilde m_N}{\Lambda}}^2\lesssim 0.1\,.
\ee

Moving to the case ii), the lightest heavy leptons can be both negatively charged or neutral. The present constraints from colliders, and particularly from the L3 experiment, only put a lower bound on the masses of the charged particles that is at $\widehat{m}_{\psi^-}\sim100\GeV$~\cite{Workman:2022ynf}, weaker than the consistency limit in Eq.~\eqref{BoundHNLMassesTheo}. The neutral particles fall in the category discussed above of the HNL and the same bounds apply also here: for the lightest masses allowed by the consistency condition, $\widehat{m}_\psi^0=500\GeV$, the corresponding constraints on the couplings of $Z$ and $W$ with a light lepton and a heavy neutral $\widehat{\psi}^0$ read
\be
\begin{gathered}
\dfrac{1}{4}\left[\dfrac{m_V+\widetilde{m}_{V'}}{\Lambda-\widetilde{M}_\psi}-\dfrac{m_V-\widetilde{m}_{V'}}{\Lambda+\widetilde{M}_\psi}\right]^2\lesssim0.1\,,\\
\left[\dfrac{\widetilde{m}_\mu \widetilde{m}_R}{\widetilde{M}_\psi^2}+\dfrac{\widetilde{m}_N}{2\widetilde{M}_\psi}\left(\dfrac{m_V+\widetilde{m}_{V'}}{\Lambda-\widetilde{M}_\psi}+\dfrac{m_V-\widetilde{m}_{V'}}{\Lambda+\widetilde{M}_\psi}\right)\right]^2\lesssim0.1\,,\qquad\qquad
\lrq{\dfrac{\widetilde{m}_R}{\widetilde{M}_\psi}}^2\lesssim0.1\,.
\end{gathered}
\label{HNLBoundsToy}
\ee
The results on the HNL apply under the assumptions of a single flavour analysis and of the specific Dirac/Majorana nature of the considered HNL that could induce a rescaling of the bounds of $1/2$ or $1/\sqrt2$ factors. These possible modifications, however, would not alter the relative relevance of these constraints and then we will stick to the bounds in Eq.~\eqref{HNLBoundsToy}. 

In the diagonalisation procedure, we have required $\Lambda\neq \widetilde{M}_\psi$. 
 In App.~\ref{APP:DEGHF} we provide the results in the degenerate limit, showing the consistency of the analysis even in this limit.

\subsubsection*{Neutrino masses}

In this simplified model, only one active neutrino gets mass, while the other two remain massless. We will then arbitrarily choose that its mass corresponds to the square root of the atmospheric mass squared difference, $\widehat{m}_\nu\sim\sqrt{\Delta m^2_\text{atm}}$.

Given the bounds on $\widetilde{m}_N/\Lambda$ discussed above,  and in particular, having identified in Eq.~\eqref{MWRequiredToy}  the range of values to explain the CDF II measurement of $M_W$,  we can now estimate the conditions to obtain the correct value for the active neutrino mass. We find a well-defined correlation between the two parameters breaking explicitly the lepton number, 
\be
\mu-15\,\epsilon\,Y_s\,v\cos\theta\approx6\eV\,,
\label{NuMassCorrelation}
\ee
obtained taking the central value for $\widetilde{m}_N/\Lambda$ in Eq.~\eqref{MWRequiredToy} and the central value for $\Delta m^2_\text{atm}$. It is then possible to extract constraints on each single parameter assuming that the other is vanishing, that at the $2\sigma$ level for $\widetilde{m}_N/\Lambda$ and $\Delta m^2_\text{atm}$ read
\be
\begin{cases}
\mu\in[4.3,\,7.5]\eV\qquad\qquad&\text{for $\epsilon=0$}\\[2mm]
\epsilon\,Y_s\in-[1.7,\,1.3]\times 10^{-12}&\text{for $\mu=0$}\,.
\end{cases}
\label{nuMassLimits}
\ee

\subsection{Relevant phenomenology at $1$-loop-level}
\label{sec:Pheno1LoopLevel}

We can move now to the analysis of the loop-level contributions, focusing on muon MDM and on the corrections to the muon mass.

\subsubsection*{Muon magnetic dipole moment}
\begin{figure}[tbh]
\centering
\includegraphics[width=0.7\textwidth]{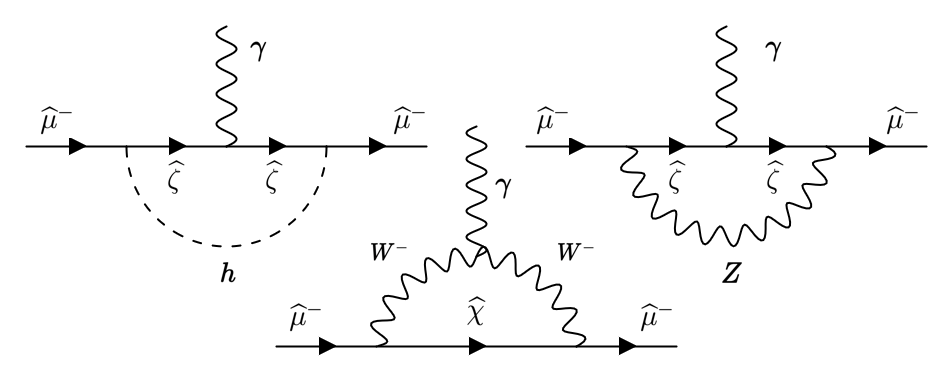}\par
\caption{\em Diagrams contributing to the $g-2$ of the muon at 1-loop in unitary gauge.}
\label{fig:g-2-diagrams}
\end{figure}

The leading EW contributions  to  $(g-2)_\mu$ are those associated with the Feynman diagrams in Fig.~\ref{fig:g-2-diagrams}, drawn in the mass basis and in the unitary gauge. 

As the internal fermion lines may be any of the neutral $\widehat{\chi}$ or charged $\widehat\zeta$ leptons defined in Eqs.~\eqref{ZetaMassEig} and \eqref{ChiMassEig}, the total EW contribution then accounts for both the SM EW part $\delta a_\mu^\text{SM-EW}$ and the NP one $\delta a_\mu^\text{NP}$,
\be
\delta a^\text{EW}_\mu\equiv \delta a_\mu^\text{SM-EW}+\delta a_\mu^\text{NP}=\delta a_\mu^h+\delta a_\mu^Z+\delta a_\mu^W\,.
\ee
Focusing on the NP term, we separately discuss the chirally enhanced and the chirally suppressed contributions: in the latter case, the contribution is suppressed by a muon mass term, besides the one due to the definition of the MDM; in the former, this additional suppression is not present. Notice that, contrary to the naive expectation, we cannot identify the chirally suppressed contributions with only those where the chirality flip is due to the muon Yukawa insertion, as muon mass terms may arise also in the interaction vertices with a scalar or gauge boson.  Thus, we may have contributions corresponding to diagrams where the chirality flip occurs due to the heavy mass in the internal lepton propagator, but  suppressed by muon mass terms in the vertices: we will consider these contributions as chirally suppressed.

We can qualitatively discuss our expectations, by looking at the couplings of the Lagrangian in Eqs.~\eqref{hLagMix}-\eqref{WLagMix}.\\

\noindent
\underline{Chirally Enhanced (CE) Contributions:}
The chirality flip has to  occur on the internal fermion line so that the total contribution is multiplied by the (heavy) mass of such a fermion. This can only occur in the diagram with the $W$ exchange.  There is no vertex $\ov{\widehat{\mu}_L}\slashed{Z}\widehat{\psi}^-_L$ and therefore no contribution is expected with a $Z$-loop and the Higgs coupling with the charged fermions $h\ov{\widehat{\mu}_L}\widehat{\psi}^-_R$ is proportional to $\widetilde m_\mu\approx\widehat{m}_\mu$; therefore this contribution would be chirally suppressed.

One concludes that there are $W$-mediated contributions involving $\widehat{N_R}$, $\widehat{S}_R$, $\widehat{\psi}^0_L$ and $\widehat{\psi}^0_R$ exchange, each of them proportional to 
\be
\widetilde{m}_N\dfrac{\widetilde{m}_R}{\widetilde{M}_\psi}\left[\dfrac{m_V+\widetilde{m}_{V'}}{\Lambda-\widetilde{M}_\psi}+\dfrac{m_V-\widetilde{m}_{V'}}{\Lambda+\widetilde{M}_\psi}\right]\times
\begin{cases}
\dfrac{\widehat{m}_{\widehat\chi}}{\Lambda} \qquad\text{for $\widehat{N_R}$ and $\widehat{S}_R$}\\[3mm]
\dfrac{\widehat{m}_{\widehat\chi}}{\widetilde{M}_\psi}\qquad\text{for $\widehat{\psi}^0_{L,R}$}
\end{cases}
\ee
As in first approximation $\widehat{m}_{\widehat{N}_R,\widehat{S}_R}\simeq\Lambda$ and $\widehat{m}_{\widehat{\psi}^0_{L,R}}\simeq\widetilde{M}_\psi$ (see Eq.~\eqref{FinalMassesNeutral}), the very last ratios are equal to $1$ and one is tempted to conclude that the sum of all those contributions would be suppressed by only two powers of the heavy fermion masses. However, this is not the case, as there is an exact cancellation between the contributions of $\widehat{N_R}$ and $\widehat{S}_R$ and the ones of $\widehat{\psi}^0_L$ and $\widehat{\psi}^0_R$. As this feature recently received special attention in Refs.~\cite{Arkani-Hamed:2021xlp,Craig:2021ksw,DelleRose:2022ygn}, we further discuss this aspect here below. 

The generic amplitude associated to the $W$-mediated diagram with an internal neutral lepton $\widehat{\chi}$ with mass $\widehat{m}_{\widehat\chi}$ reads
\begin{equation}
\begin{split}
i\mathcal{M}_{\widehat\chi}=-ig_L^2\int \dfrac{d^4k}{(2\pi)^4}\Bigg[&\ov{u}(p+q)\gamma_\mu\left(c^{\widehat\chi}_LP_L + c^{\widehat\chi}_RP_R\right)\dfrac{\slashed{k}+\widehat{m}_\chi}{k^2-\widehat{m}_\chi^2+i\epsilon}\times\\
&\times\gamma_\nu\left(c^{\widehat\chi}_{L}P_L + c^{\widehat\chi}_{R}P_R\right)\mathcal{S}^{\mu\nu}(k,p,q)u(p)\Bigg]\,,
\end{split}
\label{eq:amplitude-cancellation}
\end{equation}
where $u(p)$ ($\ov{u}(p+q)$) represents the incoming (outgoing) muon with four-momentum $p$ ($p+q$), $\mathcal{S}^{\mu\nu}(k,p,q)$ encodes the $W$-propagators and the SM interaction of the two $W$s with the photon of four-momentum $q$ and $P_{L,R}=(1\mp\gamma_5)/2$ are the chirality projectors.  Finally, $c^{\widehat\chi}_{L,R}$~\footnote{According to the initial Lagrangian, these couplings are assumed to be purely real.} are the $W$-vertices with a muon and  the normalisation of the $\widehat\chi$ is chosen so that the weak gauge coupling $g_L$ has been factorised out  in front of the integral.  The CE part of the amplitude $\mathcal{M}^{CE}_{\widehat\chi}$ corresponds to the $\widehat{m}_{\widehat\chi}$ term in the numerator of the fermion propagator,
\begin{equation}
    \begin{split}
        i\mathcal{M}^{\text{CE}}_{\widehat\chi} &= \widehat{m}_\chi c^{\widehat\chi}_L c^{\widehat\chi}_R \int \dfrac{d^4k}{(2\pi)^4} \ov{u}(p+q)\gamma_\mu\dfrac{1}{k^2-\widehat{m}_\chi^2+i\epsilon}\gamma_\nu\mathcal{S}^{\mu\nu}(k,p,q)u(p)\,,\\
        & \equiv  2\, i\,\widehat{m}_\chi\, c^{\widehat\chi}_L\, c^{\widehat\chi}_R\, \ov{u}(p+q)\, \mathcal{F}^{\widehat\chi}(p,q)\, u(p)\,,
    \end{split}
\label{eq:mass-amplitude}
\end{equation}
where the function $\mathcal{F}^{\widehat\chi}(p,q)$ is defined  so that it encodes the integration over the loop-momentum, the Lorentz structures and some factors. The leading term in the limit of $\widehat{m}_\chi\gg v$ reads
\begin{equation}
    \mathcal{F}^{\widehat\chi}(p,q) =\dfrac{e}{(16\pi^2)v^2}(\gamma_\alpha\,\slashed{q}-q_\alpha)\,\varepsilon^\alpha(q)\,,
\end{equation}
where $\varepsilon^\alpha(q)$ is the polarisation vector of the photon. 

The relevant aspect here is the absence of $\widehat{m}_\chi$ in the leading term of the function $ \mathcal{F}^{\widehat\chi}(p,q)$,  which guarantees the cancellation mentioned above. Indeed,  using the expressions for the couplings $c^{\widehat\chi}_{L,R}$ given by Eq.~\eqref{WLagMix}, 
\ba
c_L^{\widehat{N}_R} c_R^{\widehat{N}_R} &
= c_L^{\widehat{S}_R} c_R^{\widehat{S}_R} 
=-\dfrac{\tilde{m}_N \tilde{m}_R}{4\,\Lambda\,\widetilde{M}_\psi}\left[\dfrac{m_V+\widetilde{m}_{V'}}{\Lambda-\widetilde{M}_\psi}+\dfrac{m_V-\widetilde{m}_{V'}}{\Lambda+\widetilde{M}_\psi}\right]\,,
\\
c_L^{\widehat\psi^0_L} c_R^{\widehat\psi^0_L} &
=c_L^{\widehat\psi^0_R} c_R^{\widehat\psi^0_R}
=\dfrac{\tilde{m}_N \tilde{m}_R}{4\,\widetilde{M}^2_\psi}\left[\dfrac{m_V+\widetilde{m}_{V'}}{\Lambda-\widetilde{M}_\psi}+\dfrac{m_V-\widetilde{m}_{V'}}{\Lambda+\widetilde{M}_\psi}\right]\,,
\label{eq:couplings-amplitude}
\ea
we find that the total CE contribution to the amplitude reads
\begin{align}
i\mathcal{M}^\text{CE}&=
\sum_{\widehat\chi=\widehat{N}_R, \widehat{S}_R, \widehat{\psi}^0_L,\widehat{\psi}^0_R}i\mathcal{M}^\text{CE}_{\widehat\chi}\nn\\
&=-i\dfrac{\tilde{m}_N \tilde{m}_R}{2\widetilde{M}_\psi}\left[\dfrac{m_V+\widetilde{m}_{V'}}{\Lambda-\widetilde{M}_\psi}+\dfrac{m_V-\widetilde{m}_{V'}}{\Lambda+\widetilde{M}_\psi}\right]\,\ov{u}(p+q)\times
\label{g2muCE1LCancellation}
\\
&\hspace{1cm}\times\left[
\dfrac{\widehat{m}_{N_R}}{\Lambda}\mathcal{F}^{\widehat{N}_R}(p,q)
+\dfrac{\widehat{m}_{S_R}}{\Lambda}\mathcal{F}^{\widehat{S}_R}(p,q)
-\dfrac{\widehat{m}_{\psi^0_L}}{\widetilde{M}_\psi}\mathcal{F}^{\widehat{\psi}^0_L}(p,q)
-\dfrac{\widehat{m}_{\psi^0_R}}{\widetilde{M}_\psi}\mathcal{F}^{\widehat{\psi}^0_R}(p,q)
\right] u(p)\,.
\nn
\end{align}
Once taking the Leading-Order (LO) expression of the neutral heavy lepton masses as from Eq.~\eqref{FinalMassesNeutral},
\be
\widehat{m}_{N_R}=\widehat{m}_{S_R}
\simeq \Lambda\,,\qquad\qquad
\widehat{m}_{\psi^0_{L,R}}\simeq\widetilde{M}_\psi\,,
\label{LOMassesExotic}
\ee
the terms in the last line of Eq.~\eqref{g2muCE1LCancellation} sum up to zero and thus the whole CE amplitude vanishes at LO.

All in all,  at $1$-loop the CE contribution to $\delta a_\mu^\text{NP}$ arises only  at the Next-to-Leading-Order (NLO) and it is suppressed by four powers of the heavy neutral masses:
\be
  \delta a_\mu^{\text{CE-1L}}=
    \dfrac{3\,m^\text{exp}_\mu}{4\,\pi^2\,v^2}
    \dfrac{M_W^2}{\Lambda\widetilde{M}_\psi}
    \dfrac{\widetilde{m}_N\widetilde{m}_R}{\widetilde{M}_\psi}
    \left(\dfrac{m_V}{\widetilde{M}_\psi}+\dfrac{\widetilde{m}_{V'}}{\Lambda}\right)
    F_0\left(\dfrac{\Lambda^2}{M_W^2},\dfrac{\widetilde{M}_\psi^2}{M_W^2}\right)
\label{g2muCE1LNLO}    
\ee
 where the loop function is defined by
\begin{equation}
    F_0(x,y)\equiv \dfrac{3}{2}-\dfrac{x\log{y}-y\log{x}}{x-y}\,,
\end{equation}
that is negative for $x,\,y\gg 1$.\footnote{\label{FootnoteDifference} Our result differ from the one in Ref.~\cite{Arkani-Hamed:2021xlp}, that reads:
\begin{equation}
    \delta a_\mu^{\text{CE-1L}}=
    \dfrac{3\sqrt{2}\,m^\text{exp}_\mu}{4\,\pi^2\,v^2}
    \dfrac{M_W^2}{\Lambda\widetilde{M}_\psi}
    \dfrac{\widetilde{m}_N\widetilde{m}_R}{\widetilde{M}_\psi}
    \left(\dfrac{m_V}{\widetilde{M}_\psi}+\dfrac{\widetilde{m}_{V'}}{\Lambda}\right)
    F\left(\dfrac{\Lambda^2}{M_W^2},\dfrac{\widetilde{M}_\psi^2}{M_W^2}\right)\,,
\end{equation}
where the loop function is defined by
\begin{equation}
    F(x,y)\equiv \frac{x^3y\log x}{(y-x)(x-1)^3}+\frac{y^3x\log y}{(x-y)(y-1)^3}-\frac{xy(3xy-x-y-1)}{2(x-1)^2(y-1)^2}>0\,.
\end{equation}
The $\sqrt{2}$ factor follows from the different definition of the value for $v$. More relevant is the loop-function: $F_0(x,y)$ coincides with the LO expansion of the function $F(x,y)$ for $x,\,y\gg 1$,  besides the  global minus sign. Our result is consistent with the analysis in Ref.~\cite{Freitas:2014pua}.} Notice that Eq.~\eqref{g2muCE1LNLO} and the one that will follow are given in terms of the tilde quantities in order to keep the expressions more compact. 

It is worth commenting  that, although the CE contribution from $2$-loop diagrams does not present such a cancellation, it turns out to be smaller than the $20\%$ of $\delta a_\mu^{\text{CE-1L}}$ in the considered parameter space. The dominant $2$-loop CE contribution, $\delta a_\mu^{\text{CE-2L}}$, arises from a diagram similar to the one in the bottom in Fig.~\ref{fig:g-2-diagrams} with the addition of a top-bottom-loop from which the photon is emitted, see Ref.~\cite{Arkani-Hamed:2021xlp}. A rough estimation reads
\begin{equation}
    \delta a_\mu^{\text{CE-2L}}\approx - \dfrac{6\,y_t^2\, m^\text{exp}_\mu}{(16\pi^2)^2\,v^2}
    \dfrac{\widetilde{m}_N\widetilde{m}_R}{\widetilde{M}_\psi}
    \dfrac{\Lambda\widetilde{M}_\psi}{\widetilde{M}_\psi^2- \Lambda^2}
    \left(\dfrac{m_V}{\widetilde{M}_\psi}+\dfrac{\widetilde{m}_{V'}}{\Lambda}\right)
    \log{\dfrac{\widetilde{M}_\psi^2}{\Lambda^2}}\,,
\label{g2muCE2LLO}
\end{equation}
where $y_t=0.81$ is the top quark Yukawa computed at the scale $\Lambda\sim\widetilde{M}_\psi$. 

It is to be mentioned that our results agree (with the exception of what is mentioned in footnote~\ref{FootnoteDifference}) with those presented in the past literature~\cite{Kannike:2011ng,Dermisek:2013gta,Arcadi:2021cwg,Lu:2021vcp,Guedes:2022cfy,DelleRose:2022ygn} and in particular Refs.~\cite{Arkani-Hamed:2021xlp,Craig:2021ksw,DelleRose:2022ygn} focussed on the analysis of the cancellation present in the $1$-loop CE contribution.\\

\noindent
\underline{Chirally Suppressed (CS) Contributions:}
There are several CS contributions, besides the purely SM ones. We can identify a contribution from the $h$-mediated diagram, once combining the two vertices in the first line of Eq.~\eqref{hLagMix}. Moreover, a similar contribution arises from the $W$-mediated diagram with the $\widehat{\psi}^0_{L,R}$ leptons in the loop, considering the vertices in the last two lines of Eq.~\eqref{WLagMix}. In both of them, the chirality flip occurs due to a heavy lepton mass in the fermionic propagator, but the muon mass term is present in the vertex. Additional CS contributions arise again from the $W$-mediated diagram, but when the chirality flip is due to a muon Yukawa coupling in one of the muon external legs. This occurs due to the terms present in the first and last lines of Eq.~\eqref{WLagMix}, when the same vertex appears twice in the diagram.
The complete expression for the $1$-loop CS contribution reads
\begin{equation}
    \delta a_\mu^{\text{CS-1L}} =\dfrac{m^\text{exp}_\mu}{16\,\pi^2\,v^2}
    \left[-\frac{\widehat{m}_\mu \widetilde{m}_R^2}{\widetilde{M}_\psi^2}+\frac{2 m^\text{exp}_\mu}{3}\left(\left(5+2\cos2\theta_W\right)\dfrac{\widetilde{m}_R^2}{\widetilde{M}_\psi^2}-
    \dfrac{3\,\widetilde{m}_N^2}{\Lambda^2}\right)\right] \,,
\label{g2muCS1LNLO}
\end{equation} 
where we distinguish the two kind of contributions described above: the one proportional to $m^\text{exp}_\mu\widehat{m}_\mu$ comes from the diagrams with the chirality flip due to the heavy lepton masses, while the one with $(m^\text{exp}_\mu)^2$ from those with the chirality flip in the external muon legs.

If the tree-level contribution to the muon mass is the dominant one,  $\widehat{m}_\mu\simeq m_\mu^\text{exp}$,  it is possible to compute an upper bound for $\delta a_\mu^{\text{CS-1L}}$.  Fixing $\widetilde{m}_R^2/\widetilde{M}_\psi^2$ to its maximal value allowed by Eq.~\eqref{eq:exp_bounds} and with  $\widetilde{m}_N^2/\Lambda^2$ within the range given in Eq.~\eqref{MWRequiredToy} for explaining the CFM II result for $M_W$, we get $\left|\delta a_\mu^{\text{CS-1L}}\right|\lesssim 8\times 10^{-12}$, that is two orders of magnitudes smaller than necessary to explain the $(g-2)_\mu$ anomaly.

\begin{figure}[h!]
\centering
\includegraphics[width=0.48\textwidth]{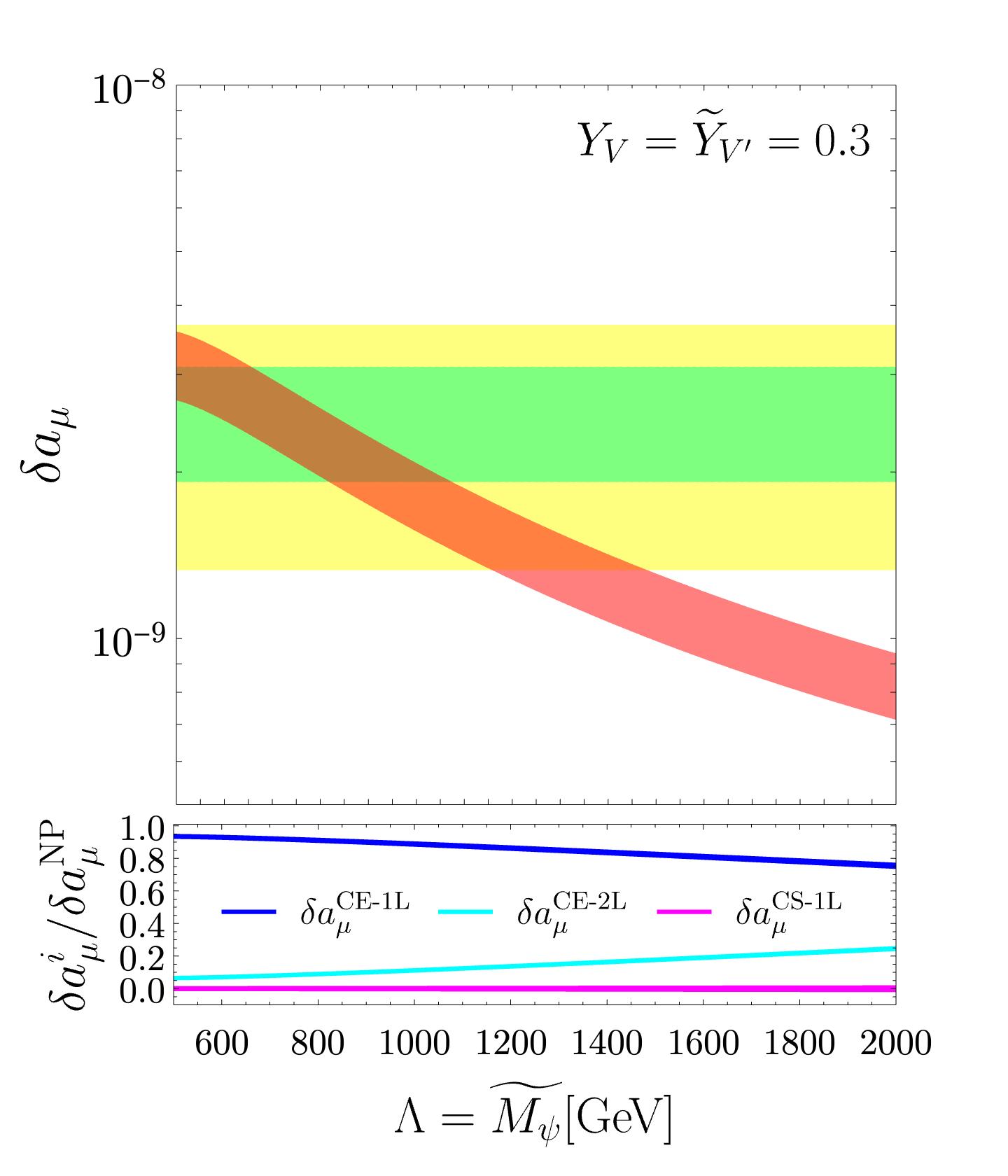}
\includegraphics[width=0.48
\textwidth]{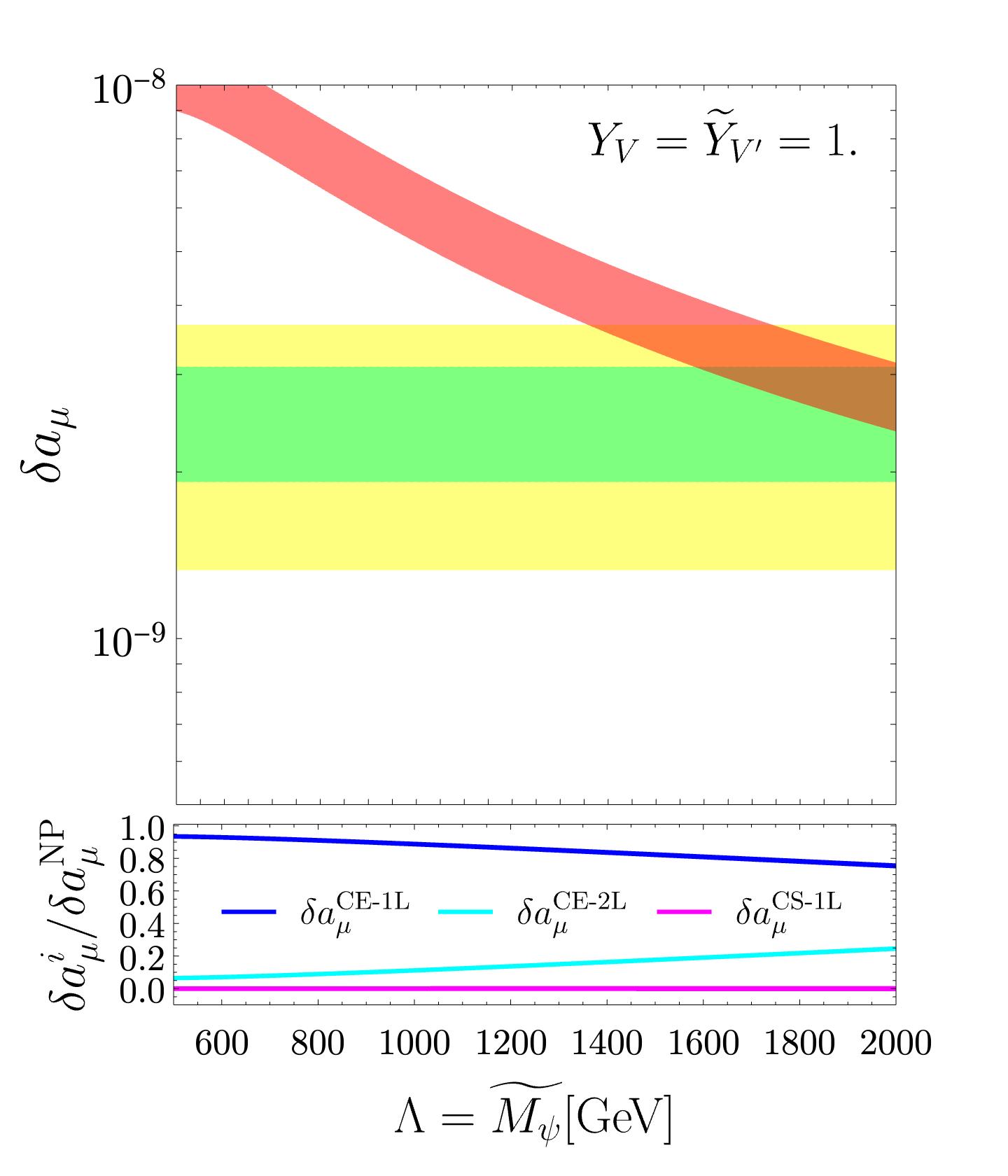}
\par
\caption{\em In the upper part of each plot, $\delta a_\mu^\text{NP}$ (red) is shown as a function of $\Lambda=\widetilde{M}_\psi$. The experimental values at $1\sigma$ ($2\sigma$) are shown in green (yellow). In the plots, $\widehat{m}_\mu=m_\mu^\text{exp}$, while $Y_V=\widetilde{Y}_{V'}$ are fixed to $0.3$ in the left plot and to $1$ in the right one. $\widetilde{m}_N/\Lambda$ can vary according to Eq.~\eqref{MWRequiredToy}, while $\widetilde{m}_R^2/\widetilde{M}_\psi^2$ is taken at its maximum value according to Eq.~\eqref{eq:exp_bounds}, together with the requirement that $|\widetilde{Y}_N|\leq1$ and $|\widetilde{Y}_R|\leq1$. In the lower part of each plot, the ratios of the different components $\delta a_\mu^{\text{CE-1L}}$ (blue), $\delta a_\mu^{\text{CE-2L}}$ (cyan) and $\delta a_\mu^{\text{CS-1L}}$ (magenta),   to the total contribution $\delta a_\mu^\text{NP}$ are shown.}
\label{fig:g-2DifferentContributions}
\end{figure}

The full NP contribution to the $(g-2)_\mu$ is given by the sum of the different terms obtained in the previous paragraphs:
\be
\delta a_\mu^\text{NP}=\delta a_\mu^{\text{CE-1L}}+\delta a_\mu^{\text{CE-2L}}+\delta a_\mu^{\text{CS-1L}}\,.
\ee
Fig.~\ref{fig:g-2DifferentContributions} shows the dependence of $\delta a_\mu^\text{NP}$ on the heavy scales assuming a simplified parameter space with $\Lambda=\widetilde{M}_\psi$ and  $\widehat{m}_\mu=m_\mu^\text{exp}$. The plots are for the effective couplings $Y_V$ and $\widetilde{Y}_{V'}$, defined as $Y_V=\sqrt2 m_V/v$ and $\widetilde{Y}_{V'}=\sqrt2 \widetilde{m}_{V'}/v$, satisfying  $Y_V=\widetilde{Y}_{V'}=0.3$ for the plot on the left and $1$ for the one on the right. In each plot, the upper part shows $\delta a_\mu^\text{NP}$ in red, while the experimental allowed region is depicted in green (yellow) at $1\sigma$ level ($2\sigma$). In the lower part, we show the ratio of each component, $\delta a_\mu^{\text{CE-1L}}$ in blue, $\delta a_\mu^{\text{CE-2L}}$ in cyan and $\delta a_\mu^{\text{CS-1L}}$ in magenta,  to the total contribution $\delta a_\mu^\text{NP}$ as a function of $\Lambda=\widetilde{M}_\psi$. The width of the curves corresponds to the range of values given in Eq.~\eqref{MWRequiredToy} within which $\widetilde{m}_N/\Lambda$ can vary. Moreover, in the whole parameter space, the condition in Eq.~\eqref{eq:exp_bounds} for $\widetilde{m}_R^2/\widetilde{M}_\psi^2$ is saturated together with the requirement that $|\widetilde{Y}_N|=\sqrt2|\widetilde{m}_N|/v$ and $|\widetilde{Y}_R|=\sqrt2|\widetilde{m}_R|/v$ are smaller than 1.

A few conclusions can be made. First of all, we can see that the CS contribution is always subdominant in the considered parameter space, as expected by having fixed $\widehat{m}_\mu=m_\mu^\text{exp}$.  The CE contribution at two loops becomes relevant, although still subdominant, only for large values of the heavy scales. Moreover, we can identify the ballpark values for $Y_V=\widetilde{Y}_{V'}$ needed to explain the $(g-2)_\mu$ anomaly at the $2\sigma$ level: on one hand, $Y_V=\widetilde{Y}_{V'}=1$ implies that the heavy masses should be as large as $2\TeV$; on the other hand, smaller values imply lower heavy scales and the requirement that $\Lambda=\widetilde{M}_\psi\gtrsim500\GeV$ implies that $Y_V=\widetilde{Y}_{V'}\gtrsim0.07$. 

\begin{figure}[h!]
\centering
\includegraphics[width=0.48\textwidth]{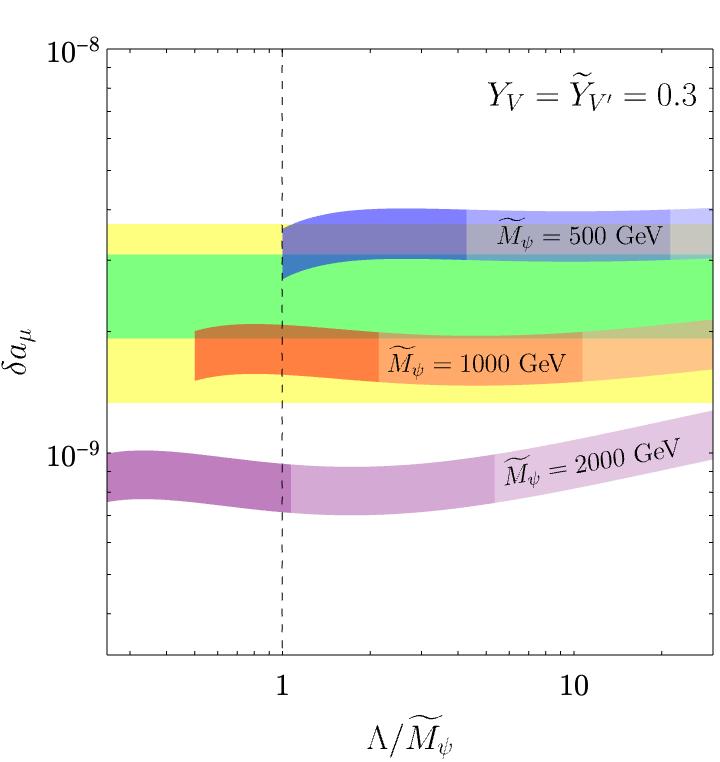}
\includegraphics[width=0.48\textwidth]{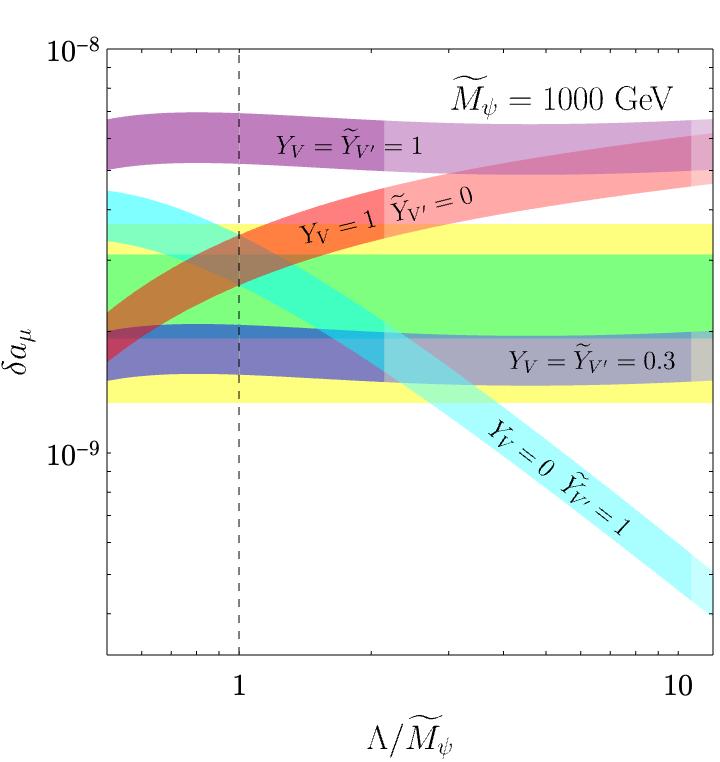}
\caption{\em $\delta a_\mu^\text{NP}$ as a function of the ratio $\Lambda/\widetilde{M}_\psi$, fixing $\widehat{m}_\mu=m_\mu^\text{exp}$. On the right, $Y_V=\widetilde{Y}_{V'}=0.3$, with $\widetilde{M}_\psi= 500\GeV$ in blue, $\widetilde{M}_\psi= 1000\GeV$ in red, $\widetilde{M}_\psi= 2000\GeV$ in purple. On the left, $\widetilde{M}_\psi=1000\GeV$, whith $Y_V$ and $\widetilde{Y}_{V'}$  fixed to different values: $Y_V=\widetilde{Y}_{V'}=1$ in purple, $Y_V=\widetilde{Y}_{V'}=0.3$ in blue, $Y_V=1$ and $\widetilde{Y}_{V'}=0$ in red, and $Y_V=0$ and $\widetilde{Y}_{V'}=1$ in cyan. Within each band, the darkest coloured (intermediate) [lightest] region is for $|\widetilde{Y}_N|<1$ ($1<|\widetilde{Y}_N|<5$) [$|\widetilde{Y}_N|>5$]. $\widetilde{m}_N/\Lambda$ can vary according to Eq.~\eqref{MWRequiredToy}, while $\widetilde{m}_R^2/\widetilde{M}_\psi^2$ saturates Eq.~\eqref{eq:exp_bounds}. The experimentally allowed values at $1\sigma$ ($2\sigma$) are shown in green (yellow). The dashed vertical line guides the eye for the ratio equal to $1$.}
\label{fig:g-2Ratio}
\end{figure}

The parameter space can also be investigated after breaking the equality between the two heavy scales and/or the relation $Y_V=\widetilde{Y}_{V'}$. In Fig.~\ref{fig:g-2Ratio}, we show  $\delta a_\mu^\text{NP}$ as a function of the ratio $\Lambda/\widetilde{M}_\psi$. In the plot on the left, we fix $Y_V=\widetilde{Y}_{V'}=0.3$ and we consider three values for $\widetilde{M}_\psi$ that span the same parameter space as  the plots in Fig.~\ref{fig:g-2DifferentContributions}: $\widetilde{M}_\psi=500\GeV$ is shown in blue, $\widetilde{M}_\psi=1000\GeV$ in red and $\widetilde{M}_\psi=2000\GeV$ in purple. The condition $\Lambda\geq500\GeV$ leads to a sharp cut on the left-hand side of the coloured regions; $\widetilde{m}_N/\Lambda$ can vary according to Eq.~\eqref{MWRequiredToy} and is responsible for the width of the bands; $\widetilde{m}_R^2/\widetilde{M}_\psi^2$ saturates  Eq.~\eqref{eq:exp_bounds}; finally, while $\widetilde{Y}_R$ is always smaller than $1$ in the whole parameter space, the opacity of the colours indicates the value of $\widetilde{Y}_N$, that is dark (intermediate) [light] for $|\widetilde{Y}_N|<1$ ($1<|\widetilde{Y}_N|<5$) [$|\widetilde{Y}_N|>5$].

The plot on the right shows complementary information. $\widetilde{M}_\psi$ is fixed to the reference value of $1000\GeV$ and the different curves show different combinations of $Y_V$ and $\widetilde{Y}_{V'}$: $Y_V=\widetilde{Y}_{V'}=1$ in purple, $Y_V=\widetilde{Y}_{V'}=0.3$ in blue, $Y_V=1$ and $\widetilde{Y}_{V'}=0$ in red and $Y_V=0$ and $\widetilde{Y}_{V'}=1$ in cyan. The same conditions on $\Lambda$, $\widetilde{m}_N/\Lambda$,  $\widetilde{m}_R^2/\widetilde{M}_\psi^2$, and $\widetilde{Y}_N$ described for the plot on the left apply also here. 

The main message following from the  two plots is that the parameter space is  large and there are many different combinations of parameters for which we can solve the muon MDM anomaly. However, the parameters need to be correlated. This feature and also the role of the constraints following from the simultaneous explanation of the $M_W$ anomaly and precision electroweak fits can be better understood  by looking at the qualitative features of the $\delta a_\mu$ contributions.  Focussing on the CE ones that are dominant in the considered parameter space,  the dependence on the six parameters ($\Lambda$, $\widetilde{M}_\psi$, $\widetilde{m}_N$, $\widetilde{m}_R$, $m_V$, $\widetilde{m}_{V'}$) is to a good approximation a dependence on their three combinations. First of all, for the CE contribution at $1$ loop, varying $\Lambda$ and $\widetilde{M}_\psi$ within $[500,\,2000]\GeV$, the function $F_0(x,\,y)$ appearing in Eq.~\eqref{g2muCE1LNLO} spans a very narrow range of value, $-[1.5,\,4]$. Taking $F_0(x,\,y)=-2.5$ in the whole parameter space, the expression in Eq.~\eqref{g2muCE1LNLO} reads:
\begin{equation}
    \delta a_\mu^{\text{CE-1L}}\simeq-\left(2\times 10^{-3}\GeV\right)
    \left[\dfrac{\widetilde{m}_N}{\Lambda}\right]
    \left[\dfrac{\widetilde{m}_R}{\widetilde{M}_\psi}\right]
    \dfrac{1}{\widetilde{M}_\psi}
    \left(\dfrac{m_V}{\widetilde{M}_\psi}+\dfrac{\widetilde{m}_{V'}}{\Lambda}\right)\,,
\label{g2muCE1LNLONaive}
\end{equation}
where the ratio in the first bracket is constrained by the CDF II measurement of $M_W$,
Eq.~\eqref{MWRequiredToy}, and the ratio in the second bracket is bounded by Eq.\eqref{eq:exp_bounds}. It follows that $\delta a_\mu^{\text{CE-1L}}$ effectively depends on only three  parameters, that is $\widetilde{M}_\psi$ and the two ratios $m_V/\widetilde{M}_\psi$ and $\widetilde{m}_{V'}/\Lambda$. We can proceed in a similar way for the CE contribution at $2$ loops and we notice that varying $\Lambda$ and $\widetilde{M}_\psi$ still within $[500,\,2000]\GeV$, the combination $\Lambda^2\log\left(\widetilde{M}_\psi^2/\Lambda^2\right)/\left(\widetilde{M}_\psi^2- \Lambda^2\right)$ spans a very  small range of values $[0.18,\,3]$.  Fixing it at $1.5$, Eq.~\eqref{g2muCE2LLO}  reads
\be    
\delta a_\mu^{\text{CE-2L}}\approx -\left(4\times 10^{-10}\GeV^{-1}\right) 
\left[\dfrac{\widetilde{m}_N}{\Lambda}\right]
\left[\dfrac{\widetilde{m}_R}{\widetilde{M}_\psi}\right]
\widetilde{M}_\psi
\left(\dfrac{m_V}{\widetilde{M}_\psi}+\dfrac{\widetilde{m}_{V'}}{\Lambda}\right)\,.
\label{g2muCE2LLONaive}
\ee

\begin{figure}[h!]
\centering
\includegraphics[width=0.58\textwidth]{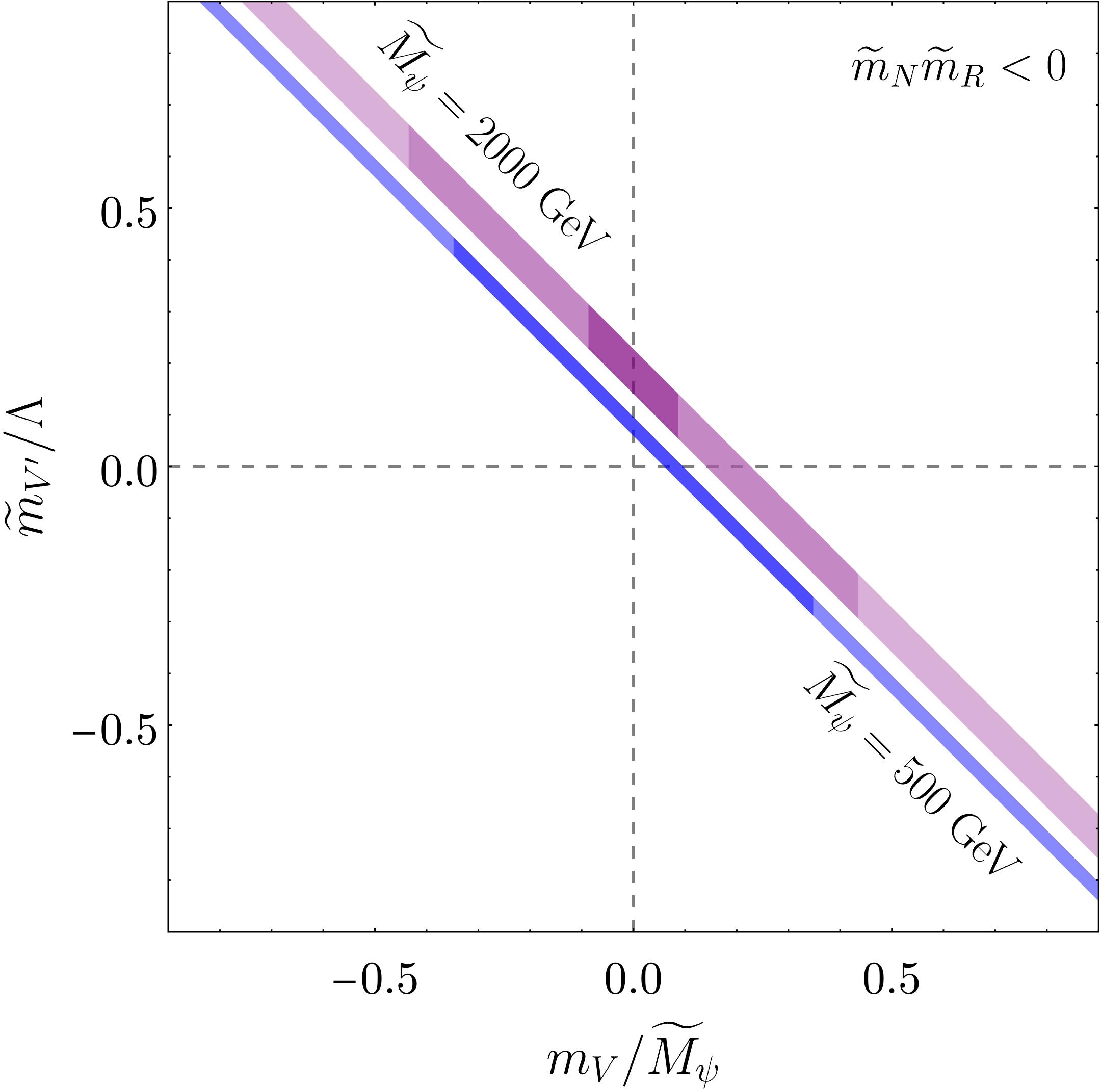}
\par
\caption{\em The parameter space $m_V/\widetilde{M}_\psi$ \vs $\widetilde{m}_{V'}/\Lambda$  which gives  $\delta a_\mu$ within its $1\sigma$ error range, for $\widetilde{m}_N\,\widetilde{m}_R<0$. The shaded regions correspond to $\widetilde{M}_\psi=500\GeV$ in blue and $\widetilde{M}_\psi=2000\GeV$ in purple. The width  of the strips is determined by $\widetilde{m}^2_N/\Lambda^2$ in the range given in Eq.~\eqref{MWRequiredToy}. For the purple band, the darkest (intermediate) [lightest] coloured region is for $|\widetilde{Y}_N|<1$ ($1<|\widetilde{Y}_N|<5$) [$|\widetilde{Y}_N|>5$], while for the blue one the dark (light) coloured region is for $|\widetilde{Y}_N|<1$ ($1<|\widetilde{Y}_N|<5$) and $|\widetilde{Y}_N|$ is never larger than $5$.}
\label{fig:3Lines}
\end{figure}

We can use these simplified expressions to easily estimate that $\delta a_\mu^{\text{CE-2L}}$ is always subdominant with respect to $\delta a_\mu^{\text{CE-1L}}$ in the whole considered parameter space $\widetilde{M}_\psi$ \vs $\Lambda$. Moreover,  taking $\widetilde{m}^2_N/\Lambda^2$ as in Eq.~\eqref{MWRequiredToy} and $\widetilde{m}^2_R/\widetilde{M}^2_\psi$ at its largest value in Eq.~\eqref{eq:exp_bounds}, for fixed values of $\widetilde{M}_\psi$, the region where the experimental measurement for the muon MDM is reproduced reduces to an anti-diagonal straight strip in the parameter space $m_V/\widetilde{M}_\psi$ \vs $\widetilde{m}_{V'}/\Lambda$. Changing the value of $\widetilde{M}_\psi$ simply translates into moving this strip in the plane, maintaining however the slop. This is shown in the plot in Fig.~\ref{fig:3Lines}. The two shaded regions represent different values of $\widetilde{M}_\psi$, that is $500\GeV$ in blue and $2000\GeV$ in purple, such that $\delta a_\mu$ matches the experimental measurement at $1\sigma$. The intensity of the colours indicate the value of the Yukawa coupling $Y_V$, which is smaller than $1$ for the darkest colour, in the range $[1,\,5]$ for the intermediate colour, and larger than $5$ for the lightest colour: notice that for $\widetilde{M}_\psi=500\GeV$ and $2000\GeV$, $|\widetilde{Y}_N|$ is never larger than $5$ in the shown parameter space. The width of the strips corresponds to $\widetilde{m}_N/\Lambda$ varying in the range given by Eq.~\eqref{MWRequiredToy}. Commenting on the signs of $\widetilde{m}_N$ and $\widetilde{m}_R$, the two blue and purple strips correspond to the case with opposite defined signs for the two parameters; instead, if the product $\widetilde{m}_N\,\widetilde{m}_R$ would be positive, then the two coloured strips would appear in the bottom-left part of the plane, symmetric with respect to the origin of the coordinate system. 

Notice that, as only an upper bound on $\widetilde{m}_R/\widetilde{M}_\psi$ is provided by the current data, one may consider smaller values  of this ratio. This would imply wider regions in the  $m_V/\widetilde{M}_\psi$ \vs $\widetilde{m}_{V'}/\Lambda$ parameter space: for fixed values of $\widetilde{M}_\psi$, reducing $|\widetilde{m}_R/\widetilde{M}_\psi|$ translates into larger values for $|Y_V|$ and $|Y_{V'}|$, that can possibly be restricted by the perturbativity requirement on these Yukawa couplings. 

If in the future, deviations from the SM model predictions will be found such that a lower bound for $|\widetilde{m}_R/\widetilde{M}_\psi|$ can be fixed, then the correlation between the three (combination of) parameters entering Fig.~\ref{fig:3Lines} will be uniquely determined.  This is  possible in this model because $\widetilde{m}^2_N/\Lambda^2$ spans a reduced range of values in order to reproduce the CDF II measurement on $M_W$.

\subsubsection*{Muon mass}
The diagrams in Fig.~\ref{fig:g-2-diagrams},  after removing the photon leg, represent contributions to the muon mass at $1$-loop. The absence of the photon, however,  has an important consequence: the cancellation present in the CE contribution at $1$-loop to $\delta a_\mu^\text{NP}$ is now absent. Indeed,  the muon mass correction reads exactly as Eq.~\eqref{eq:amplitude-cancellation}, but with the $\mathcal{S}^{\mu\nu}(k,p)$ function that now encodes only the $W$-propagator. The amplitude is divergent and dependent on the heavy-fermion mass in the loop.  Writing the amplitude as in Eq.~\eqref{eq:mass-amplitude}, we  have to replace the $\mathcal{F}^{\widehat{\chi}}(p,q)$ function by a new one $\mathcal{G}^{\widehat{\chi}}(p)$ that reads
\begin{equation}
   \mathcal{G}^{\widehat{\chi}}(p)=- \dfrac{1}{(16\pi^2)}\dfrac{\widehat{m}_\chi^2}{v^2}\left[1+\log\left(\dfrac{\mu^2_\text{Ren}}{\widehat{m}_\chi^2}\right)\right]\,,
\label{GFunction}
\end{equation}
valid in the limit of $\widehat{m}_\chi\gg v$, where $\mu_\text{Ren}$ stands for the renormalisation scale in the $\ov{MS}$ scheme.~\footnote{As our model is renormalisable, $\mu_\text{Ren}$ is the renormalisation scale. On the contrary, $\mu_\text{Ren}$ represents the physical cut-off in Ref.~\cite{Arkani-Hamed:2021xlp}.}

Summing up all the terms as in Eq.~\eqref{g2muCE1LCancellation}, no cancellation, in general, can take place due to the presence of the mass dependence and the log in the $\mathcal{G}^{\widehat{\chi}}(p)$ functions. The final result for the dominant $1$-loop contribution at the LO to the muon mass reads
\begin{equation}
    \delta m_\mu= 
    -\dfrac{\widetilde{m}_N\,\widetilde{m}_R\,\Lambda}{8\,\pi^2\,v^2}
    \left(\dfrac{m_V}{\widetilde{M}_\psi}+\dfrac{\widetilde{m}_{V'}}{\Lambda}\right)
    \left[1+\dfrac{1}{\widetilde{M}_\psi^2-\Lambda^2}\left(\widetilde{M}^2_\psi
    \log{\dfrac{\mu_\text{Ren}^2}{\widetilde{M}_\psi^2}}-
    \Lambda^2
    \log{\dfrac{\mu_\text{Ren}^2}{\Lambda^2}}\right)\right]\,,
\label{1loopMmu}
\end{equation}

\begin{figure}[tbh]
\centering
\includegraphics[width=0.48\textwidth]{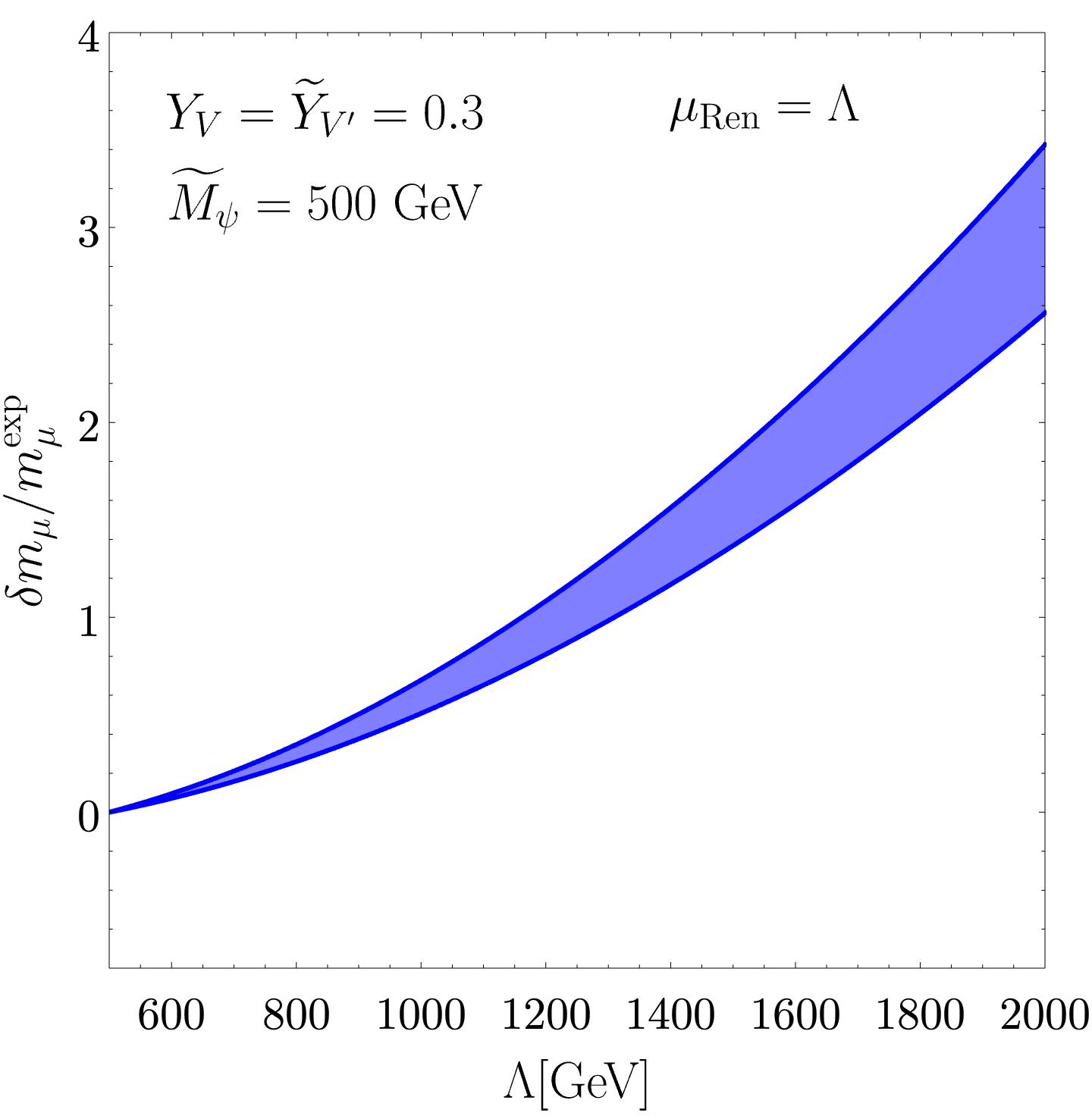}
\includegraphics[width=0.48\textwidth]{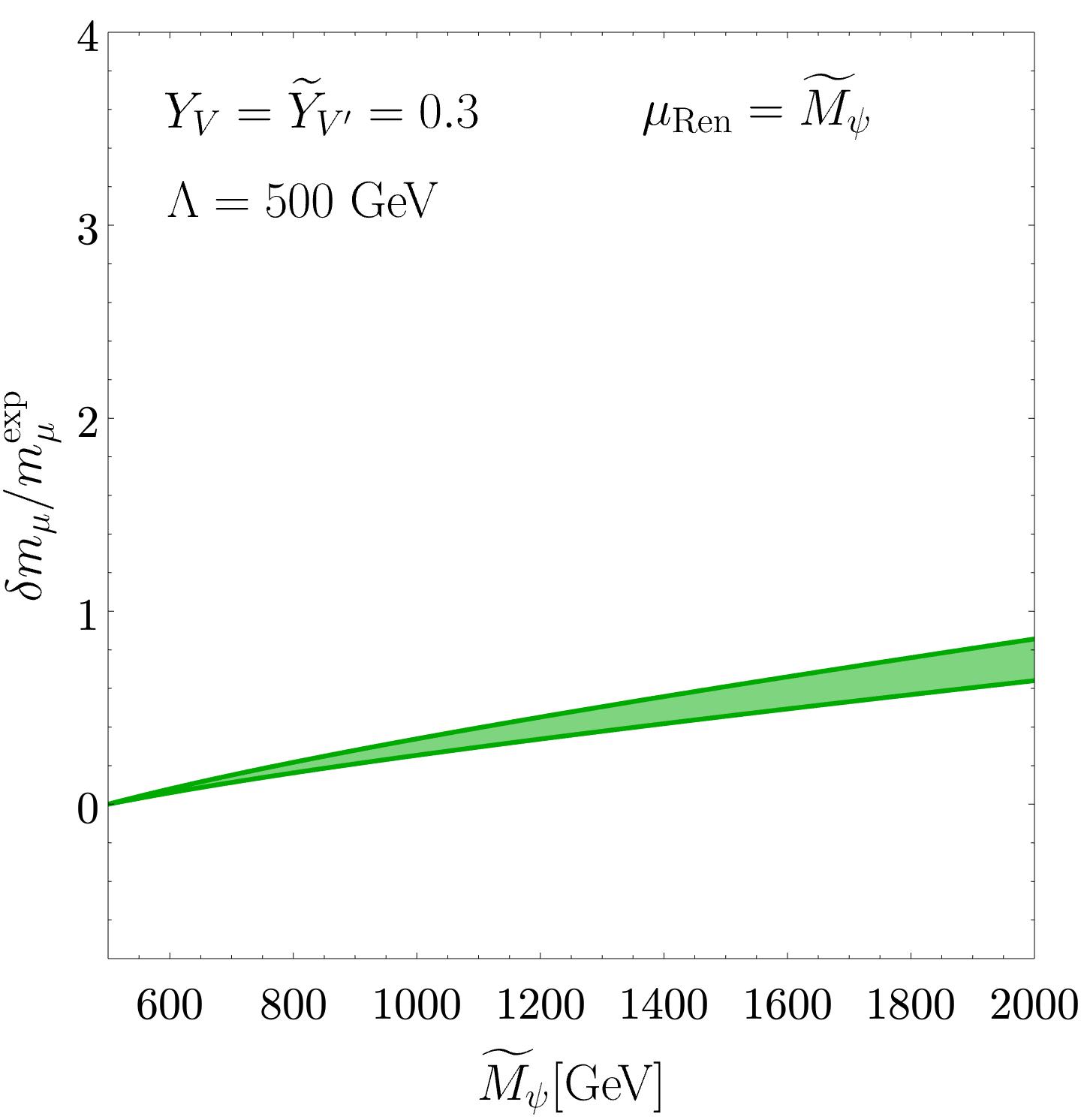}
\caption{\em Dependence of $\delta m_\mu$ on $\Lambda$ ($\widetilde{M}_\psi$) on the left (right), fixing $\widetilde{M}_\psi=500\GeV$ ($\Lambda=500\GeV$) and the running scale $\mu_\text{Ren}=\Lambda$ ($\mu_\text{Ren}=\widetilde{M}_\psi$). $Y_V$ and $\widetilde{Y}_{V'}$ have been taken both equal to $0.3$. The width of the bands corresponds to $\widetilde{m}^2_N/\Lambda^2$ within the range in Eq.~\eqref{MWRequiredToy}. $\widetilde{m}^2_R/\widetilde{M}^2_\psi$ has been fixed to the largest value in Eq.~\eqref{eq:exp_bounds} and $\widetilde{m}_N\widetilde{m}_R<0$.}
\label{fig:muonmass}
\end{figure}

The left plots in Fig.~\ref{fig:muonmass} estimates the dependence of $\delta m_\mu/m_\mu^\text{exp}$ on $\Lambda$ having fixed $\widetilde{M}_\psi=500\GeV$ and the running scale such that $\mu_\text{Ren}=\Lambda$. In the right plot, the role of $\Lambda$ and $\widetilde{M}_\psi$ are interchanged. As we can see, for the considered simplified choice of the parameters, $|\delta m_\mu|$ increases and, for $\widetilde{M}_\psi=500\GeV$, it can be even larger than the experimental value of the muon mass, at larger values of the masses of the heavy leptons. Choosing $\mu_\text{Ren}$ to coincide with the largest scale among $\widetilde{M}_\psi$ and $\Lambda$ reduces the impact of the higher loop-level contributions. 

Interestingly, in the limit $\mu_\text{Ren}=\widetilde{M}_\psi=\Lambda$, the term in the squared brackets vanishes at leading order: Eq.~\eqref{GFunction} trivially shows that the log-dependent term is identically zero as far as the renormalisation scale coincides with the mass of the exotic fermion; however, even the constant term does not lead to any contributions at leading order once $\widetilde{M}_\psi=\Lambda$ and the reason can be understood repeating the analysis done for the muon $g-2$ in Eqs.~\eqref{eq:mass-amplitude}-\eqref{LOMassesExotic}, substituting  $\mathcal{F}^{\widehat\chi}(p,q)$ with $\mathcal{G}^{\widehat{\chi}}(p)$. This can be more explicitly appreciated considering the gauge boson couplings to fermions in the limit $\widetilde{M}_\psi=\Lambda$, reported in App.~\ref{APP:DEGHF}: at the expansion order considered, the singlet and the doublet exotic fields couples to the elementary muon fields exactly in the same way, except for a global sign.

We can simplify the expression in Eq.~\eqref{1loopMmu} in the same line as we did for the muon MDM: varying $\Lambda$ and $\widetilde{M}_\psi$ within $[500,\,2000]\GeV$ and taking the renormalisation scale equal to the largest among $\Lambda$ and $\widetilde{M}_\psi$, we get
\be 
\delta m_\mu\simeq -[0,0.7]\times
\left[\dfrac{\widetilde{m}_N}{\Lambda}\right]
\left[\dfrac{\widetilde{m}_R}{\widetilde{M}_\psi}\right]
\widetilde{M}_\psi
\left(\dfrac{m_V}{\widetilde{M}_\psi}+\dfrac{\widetilde{m}_{V'}}{\Lambda}\right)\,.
\label{1loopMmuNaive}
\ee

Given the similarity of the expressions for the CE contributions of the $\delta a_\mu$ and $\delta m_\mu$, it is useful to combine $\delta m_\mu$ and $\delta a_\mu^\text{CE}$ in a single formula, using the exact expressions in Eqs.~\eqref{g2muCE1LNLO}, \eqref{g2muCE2LLO} and \eqref{1loopMmu}, we get
\be
\begin{aligned}
\dfrac{\delta a_\mu^{\text{CE-1L}}}{\delta m_\mu}&=-
    \dfrac{6\,m_\mu^\text{exp}M_W^2}{\Lambda^2\widetilde{M}^2_\psi}F_0\left(\dfrac{\Lambda^2}{M_W^2},\dfrac{\widetilde{M}_\psi^2}{M_W^2}\right)
    \left[1+\dfrac{1}{\widetilde{M}_\psi^2-\Lambda^2}\left(\widetilde{M}^2_\psi\log{\dfrac{\mu_\text{Ren}^2}{\widetilde{M}_\psi^2}}-
    \Lambda^2\log{\dfrac{\mu_\text{Ren}^2}{\Lambda^2}}\right)\right]^{-1}\\
\dfrac{\delta a_\mu^{\text{CE-2L}}}{\delta m_\mu}&=
    \dfrac{3\,y_t^2}{16\pi^2}\, \dfrac{m_\mu^\text{exp}}{\widetilde{M}_\psi^2-\Lambda^2}\log{\dfrac{\widetilde{M}_\psi^2}{\Lambda^2}}\left[1+\dfrac{1}{\widetilde{M}_\psi^2-\Lambda^2}\left(\widetilde{M}^2_\psi\log{\dfrac{\mu_\text{Ren}^2}{\widetilde{M}_\psi^2}}-
    \Lambda^2\log{\dfrac{\mu_\text{Ren}^2}{\Lambda^2}}\right)\right]^{-1}\,,
\end{aligned}
\label{CorrelationdeltaamuMmu}
\ee
that hold whenever $\delta m_\mu\neq0$. When instead $\delta m_\mu$ vanishes, that is whenever $\mu_\text{Ren}=\widetilde{M}_\psi=\Lambda$, there is no correlation between these two quantities and $\delta a_\mu$ can fit the corresponding experimental value although no quantum correction to the muon mass is present at $1$-loop and at LO.

\begin{figure}[h!]
\centering
\includegraphics[width=0.58\textwidth]{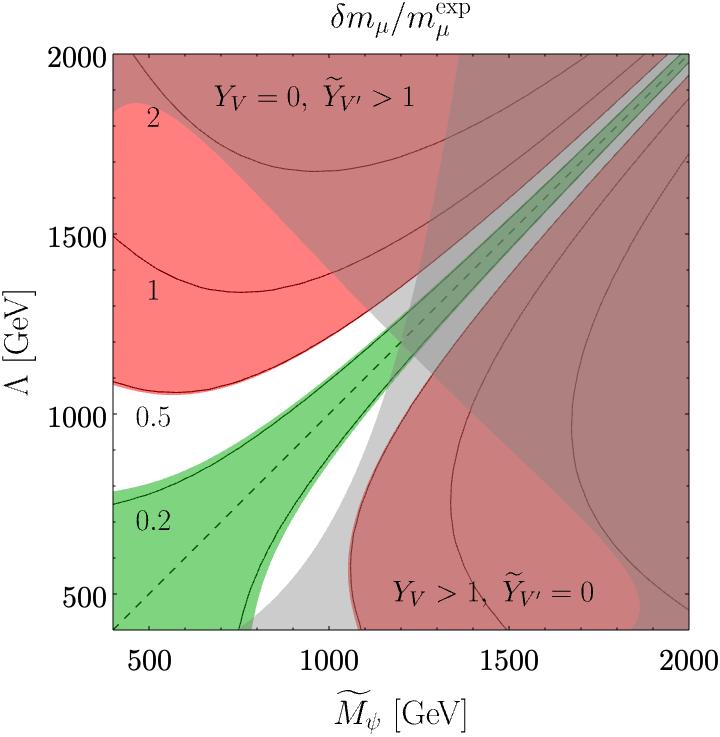}
\caption{\em $\delta m_\mu/m_\mu^{\exp}$ in the parameter space of $\widetilde{M}_\psi$ \vs $\Lambda$. The black curves represent the value of the ratio $\delta m_\mu/m_\mu^{\exp}$, with the dashed one referring to the case when it is vanishing. In the whole parameter space, $\delta a_\mu$ coincides with the corresponding experimental central value. The green region is when the loop-level contribution to the muon mass is smaller than the tree-level one by the $30\%$, while the white one is when it is larger than the $30\%$ but still smaller than the tree-level one. Finally, the red region, instead, is when the loop-level contribution is larger than the tree-level one. The two grey regions provide an intuition of when $|Y_V|$ and $|\widetilde{Y}_{V'}|$ get larger than $1$, but still smaller than $5$.}
\label{fig:deltamuvsmuexp}
\end{figure}

Fig.~\ref{fig:deltamuvsmuexp} illustrates the parameter space of $\widetilde{M}_\psi$ \vs $\Lambda$ when $\delta a_\mu$ coincides with the corresponding experimental central value, with the black curves showing the values of the ratio $\delta m_\mu/m_\mu^{\exp}$. The plot has been obtained using the complete expressions in Eqs.~\eqref{g2muCE1LNLO}, \eqref{g2muCE2LLO} and \eqref{1loopMmu}, fixing the renormalisation scale $\mu_\text{Ren}$ equal to the largest among $\Lambda$ and $\widetilde{M}_\psi$. The colours describe the relationship between the tree-level and the loop-level contributions to the muon mass: the loop contribution is larger than the tree-level one in the red area, while it is smaller than the $50\%$ ($30\%$) of the tree-level one in the  white (green) one. To draw these conditions, in each point of the parameter space, the value of the tree-level contribution is fixed by the requirement that
\be
\widehat{m}_\mu=m_\mu^\text{(2\TeV)}-\delta m_\mu\,,
\ee
where $m_\mu^\text{(2\TeV)}=103.62\MeV$ is the value of the muon mass at $2\TeV$, obtained performing the RG running from the $M_Z$ scale up to $2\TeV$ within the SM (see Ref.~\cite{Huang:2020hdv} for the values at the scale $M_Z$ and for example Ref.~\cite{Arason:1991ic} for the RG running within the SM). In grey, we indicate the regions where $1<|Y_V|<5$ with $\widetilde{Y}_{V'}=0$, or  $1<|\widetilde{Y}_{V'}|<5$ with $Y_V=0$.

All in all, in the whole considered parameter space, the correct value of the muon $(g-2)$ is achieved without the necessity of an unnaturally large cancellation between tree- and loop-level contributions to the muon mass.

\section{Discussion on the three-generation extension}
\label{sec:ThreeGen}

This section discusses the main features of considering the three generations of leptons as in the SM. 

A very economical solution in terms of number of both fields and parameters would still consider the NP part of the spectrum as in the one-generation model considered in the previous sections. Not introducing any additional symmetry, the couplings between exotic and the SM fields are in general promoted to vectors and matrices in the flavour space: in particular, $Y_e$ would be a $3\times3$ matrix, while $Y_N$, $Y_S$, $Y_R$ and $M_L$ would be tridimensional vectors. The attractive aspect of this setup is the possibility to correctly describe the pattern of lepton masses and mixings: as first shown in Ref.~\cite{Gavela:2009cd}, with only one $N_R$ and one $S_R$, it is possible to uniquely determine the structure of the Dirac Yukawa vectors in order to describe the PMNS mixing matrix and a neutrino spectrum (with both mass ordering) with the lightest neutrino being massless. 
As now also the electron couplings would be modified, the LFU ratios, which impose strong bounds in the one-generation model, would simply lead to the condition that electron and muon couplings to the $W$ gauge boson should be very similar to each other. The solution to explain the CDF II measurement of $M_W$ would be in the same ballpark as in the one-generation model -- actually, it would change by a factor $2$.

Regarding the computations for the MDMs, no relevant change is expected, as indeed the cancellation in the CE contribution at LO would still occur and therefore the associated phenomenology discussed in the one-generation model would still hold. 

The main drawback of this simple and elegant scenario is the presence of flavour-changing neutral currents. The exotic fermions would be flavour blind and therefore the same diagrams that contribute to the lepton MDM would also contribute to the radiative lepton decay, with the same dependence on the parameters -- the cancellation occurring in the 1L-CE contributions would also occur for this flavour changing processes. The net result is that $\mu\to e\gamma$ would completely exclude the parameter space interesting for the $\delta a_\mu$ solution.\\

A second possibility is that the NP part of the spectrum is extended so that  there are three replicas of the fields considered in the one-generation construction.  In this scenario the number of parameters would be largely increased as most of the parameters appearing in Eq.~\eqref{1GenLag} would be promoted to be $3\times3$ matrices in the flavour space. The expectation is that all the observables may be fitted but at the price of a very weak predictive power.

The latter possibility is certainly not economical in terms of number of fields and parameters.  A radical improvement in predictivity can be obtained  e.g. by implementing the family lepton number as a good symmetry of the Lagrangian, broken by the Majorana terms that may also be responsible for introducing the lepton mixings. Each exotic generation would then interact only with one SM lepton generation  and the Lagrangian in Eq.~\eqref{1GenLag} would simply be repeated three times. Each sector would have its own parameters and therefore the possibility of correlations between observables involving different flavours is strongly unlikely. The bottom line is that the three sectors could then be treated independently.   A less radical but  very interesting possibility is to extend the horizontal symmetries invoked to explain the pattern of fermion masses and mixings
to the NP part of the spectrum.
Thus,  there exist a variety of potential  generalisations to three generations to explore.

\section{Concluding remarks}
\label{sec:Conc}

The new measurement of the $W$ gauge boson mass from the CDF II collaboration, if confirmed, is still another indication of physics beyond the Standard Model. It is an intriguing  possibility to explain this deviation in frameworks that describe massive light active neutrinos: this is the case of the Low-Scale Seesaw constructions where the sterile lepton species may live at the TeV scale and then possibly be produced and detected at colliders. Even more fascinating would be to explain within the same framework a long-standing anomaly typically associated with low-energy physics, that is the tension between the theoretical prediction and the experimental determination of the muon anomalous magnetic moment. In this paper, we provide a proof of concept that such a construction can indeed be realised. 

We focus on a renormalisable one-generation scenario, extending the Standard Model spectrum with two additional sterile species and one pair of vector-like lepton $SU(2)_L$-doublets that interact only with the muon and the muonic neutrino.  We have studied the parameter space of the model pointing out that we can solve the $M_W$ and $(g-2)_\mu$ anomalies at the $2\sigma$ level together with reproducing the light active neutrino mass scale, without any relevant fine-tuning. This is achieved for exotic lepton masses in the range $[0.5,\,2]\TeV$, smaller than the scale naively expected in the effective field theory description. This is due to an accidental cancellation occurring between different contributions at $1$-loop to the $(g-2)_\mu$. This cancellation has been discussed in the lepton flavour basis in Refs.~\cite{Arkani-Hamed:2021xlp,Craig:2021ksw,DelleRose:2022ygn}, with the aim of understanding if an underlying explanation would be present. We have analysed it instead in the lepton mass basis and we concluded that it simply follows from the peculiarity of the couplings of the muon with the exotic states -- see Eq.~\eqref{eq:couplings-amplitude}: this is the direct consequence of the chosen spectrum and of the Standard Model gauge symmetry invariance. With respect to the past literature, besides the analysis of the $(g-2)_\mu$ discussed above, focusing in a different part of the parameter space, we explain for the first time in this context the lightness of the active neutrinos and the new measurement from the CDF II collaboration of the $M_W$ mass.

We  have also discussed the possible extensions to the three-generation case. The minimal  scenario,  without any additional new fields in the spectrum beyond the ones already considered,  preserves the positive features of the one-generation construction  and  avoids the strong bounds from the observed lepton flavour universality (see the ratios in Eq.~\eqref{LFVratiosDefinitions}).  However, as the exotic fields are flavour-blind, radiative lepton decays rule it out. A realistic three-generation model requires an extension of the exotic spectrum.  An interesting possibility to explore is to impose a flavour symmetry on the spectrum so that the predictivity of such models is preserved.  The appeal of such frameworks would also be associated with the future direct searches of new physics at colliders, where the exotic leptons may be produced and detected, confirming or ruling them out.

\section*{Acknowledgements}
The authors are grateful to E.~Fern\'andez-Mart\'inez, F.~Feruglio, G.~Guedes, K. Harigaya and D.M.~Straub for useful discussions. A.d.G. and L.M. thanks the Institute of Theoretical Physics and the Faculty of Physics of the University of Warsaw for hospitality during the development of this project. S.P. thanks the Institute of Theoretical Physics of the Universidad Autónoma de Madrid for hospitality during the development of this project.
A.d.G. and L.M. acknowledge partial financial support by the Spanish Research Agency (Agencia Estatal de Investigaci\'on) through the grant IFT Centro de Excelencia Severo Ochoa No CEX2020-001007-S and by the grant PID2019-108892RB-I00 funded by MCIN/AEI/ 10.13039/501100011033, by the European Union's Horizon 2020 research and innovation programme under the Marie Sk\l odowska-Curie grant agreement No 860881-HIDDeN. 
The research of S.P. has received partial financial support by the Polish Science Centre (NCN),   grant DEC-2016/23/G/ST2/04301

\begin{appendices}
\section{Degenerate heavy fermions}
\label{APP:DEGHF}
In the limit $\Lambda \approx \widetilde{M}_\psi$ the heavy fermions are almost degenerate. In such case, many of the couplings presented in the main discussion seem to suffer from a divergence of the type $(\Lambda-\widetilde{M}_\psi)^{-1}$. Such divergence is an artifact and the corresponding case must be treated separately to cure it. In such section, we show how the interaction Lagrangian of Eq.s~\eqref{hLagMix}, \eqref{ZLagMix}, and \eqref{WLagMix} looks like in such limit. Such results must be used whenever $v/|\Lambda-\widetilde{M}_\psi| \gtrsim 1$.

As this assumption has consequences only on the neutral sector, we will focus on it. The masses are now given by
\ba
\widehat{m}_\nu &=\dfrac{\mu\,\widetilde{m}_N^2}{\Lambda^2}-\dfrac{2\,\epsilon\,\widetilde{m}_N\, m_S \cos{\theta}}{\Lambda} \,,\\
\widehat{m}_{N_R} &= \Lambda+\dfrac{m_V+\widetilde{m}_{V'}}{2}\,,\\
\widehat{m}_{S_R}&= \Lambda-\dfrac{m_V+\widetilde{m}_{V'}}{2}\,,\\
\widehat{m}_{\psi_L^0} &= \Lambda-\dfrac{m_V+\widetilde{m}_{V'}}{2}\,,\\
\widehat{m}_{\psi_R^0} &= \Lambda+\dfrac{m_V+\widetilde{m}_{V'}}{2}\,,\\
\label{FinalMassesNeutral-deg}
\ea
where we have shown the leading and next-to-leading order for each mass. As can be seen, the splitting between the masses increases and now depends on $m_V$ and $\widetilde{m}_{V'}$.

Again, as this phenomenology arises only in the neutral sector, we will omit interactions among two charged fields. The mixing degenerate Lagrangian then reads
\begin{align}
\sL_h^\text{deg}  \supset & -\dfrac{h}{v}
\left\{-\dfrac{\widetilde{m}_N}{2}\,
\ov{\widehat\nu_L}\dfrac{\widehat{N}_R+i\widehat{S}_R-\widehat{\psi}^{0c}_L-i\widehat{\psi}_R^{0}}{2}+\text{h.c.}\right\}\,,\\
\hspace{2cm}
\sL_Z^\text{deg} \supset & -\dfrac{\sqrt{g_L^2+g_Y^2}}{2}Z_\mu\left\{-\dfrac{\widetilde{m}_N}{\Lambda}\,\ov{\widehat{\nu}_L}\gamma^\mu\dfrac{ \widehat{N}_R^c+i\widehat{S}_R^c-\widehat{\psi}^0_L-i\widehat{\psi^{0c}_R}}{2}+\text{h.c.}\right\}\,,\\
\hspace{2cm}\sL_W^\text{deg}  \supset & -\dfrac{g_L}{\sqrt{2}}W^-_\mu
\left\{-\dfrac{\widetilde{m}_N}{\Lambda}\,\ov{\mu_L}\gamma^\mu\dfrac{\widehat{N}_R^c+i\widehat{S}_R^c-\widehat{\psi}_L^0-i\widehat{\psi}_R^{0c}}{2} \right.\\
&\hspace{2cm}\nonumber+\left.\dfrac{\widetilde{m}_R}{\Lambda}\,\ov{\widehat\mu_R}\gamma^\mu\dfrac{\widehat{N}_R-i\widehat{S}_R+\widehat{\psi}_L^{0c}-i\widehat{\psi}_R^{0}}{2}\right\}\,,
\end{align}
where we have written only the leading-order terms for each coupling.


\end{appendices}

\footnotesize

\bibliography{bibliography}{}

\providecommand{\href}[2]{#2}\begingroup\raggedright\begin{thebibliography}{10}

\bibitem{ATLAS:2012yve}
{\bf ATLAS} Collaboration, G.~Aad {\em et.~al.}, {\it {Observation of a new
  particle in the search for the Standard Model Higgs boson with the ATLAS
  detector at the LHC}},  Phys. Lett. B {\bf 716} (2012) 1--29,
  [\href{http://arxiv.org/abs/1207.7214}{{\tt arXiv:1207.7214}}].

\bibitem{CMS:2012qbp}
{\bf CMS} Collaboration, S.~Chatrchyan {\em et.~al.}, {\it {Observation of a
  New Boson at a Mass of 125 GeV with the CMS Experiment at the LHC}},  Phys.
  Lett. B {\bf 716} (2012) 30--61, [\href{http://arxiv.org/abs/1207.7235}{{\tt
  arXiv:1207.7235}}].

\bibitem{CDF:2022hxs}
{\bf CDF} Collaboration, T.~Aaltonen {\em et.~al.}, {\it {High-precision
  measurement of the W boson mass with the CDF II detector}},  Science {\bf
  376} (2022), no.~6589 170--176.

\bibitem{Kersten:2007vk}
J.~Kersten and A.~Y. Smirnov, {\it {Right-Handed Neutrinos at CERN LHC and the
  Mechanism of Neutrino Mass Generation}},  Phys. Rev. D {\bf 76} (2007)
  073005, [\href{http://arxiv.org/abs/0705.3221}{{\tt arXiv:0705.3221}}].

\bibitem{Abada:2007ux}
A.~Abada, C.~Biggio, F.~Bonnet, M.~B. Gavela, and T.~Hambye, {\it {Low energy
  effects of neutrino masses}},  JHEP {\bf 12} (2007) 061,
  [\href{http://arxiv.org/abs/0707.4058}{{\tt arXiv:0707.4058}}].

\bibitem{Blennow:2022yfm}
M.~Blennow, P.~Coloma, E.~Fernandez-Martinez, and M.~Gonzalez-Lopez, {\it
  {Right-handed neutrinos and the CDF II anomaly}},  Phys. Rev. D {\bf 106}
  (2022), no.~7 073005, [\href{http://arxiv.org/abs/2204.04559}{{\tt
  arXiv:2204.04559}}].

\bibitem{Arias-Aragon:2022ats}
F.~Arias-Aragon, E.~Fernandez-Martinez, M.~Gonzalez-Lopez, and L.~Merlo, {\it
  {Dynamical Minimal Flavour Violating inverse seesaw}},  JHEP {\bf 09} (2022)
  210, [\href{http://arxiv.org/abs/2204.04672}{{\tt arXiv:2204.04672}}].

\bibitem{Wyler:1982dd}
D.~Wyler and L.~Wolfenstein, {\it {Massless Neutrinos in Left-Right Symmetric
  Models}},  Nucl. Phys. B {\bf 218} (1983) 205--214.

\bibitem{Mohapatra:1986bd}
R.~N. Mohapatra and J.~W.~F. Valle, {\it {Neutrino Mass and Baryon Number
  Nonconservation in Superstring Models}},  Phys. Rev. D {\bf 34} (1986) 1642.

\bibitem{Bernabeu:1987gr}
J.~Bernabeu, A.~Santamaria, J.~Vidal, A.~Mendez, and J.~W.~F. Valle, {\it
  {Lepton Flavor Nonconservation at High-Energies in a Superstring Inspired
  Standard Model}},  Phys. Lett. B {\bf 187} (1987) 303--308.

\bibitem{Malinsky:2005bi}
M.~Malinsky, J.~C. Romao, and J.~W.~F. Valle, {\it {Novel supersymmetric SO(10)
  seesaw mechanism}},  Phys. Rev. Lett. {\bf 95} (2005) 161801,
  [\href{http://arxiv.org/abs/hep-ph/0506296}{{\tt hep-ph/0506296}}].

\bibitem{Esteban:2020cvm}
I.~Esteban, M.~C. Gonzalez-Garcia, M.~Maltoni, T.~Schwetz, and A.~Zhou, {\it
  {The fate of hints: updated global analysis of three-flavor neutrino
  oscillations}},  JHEP {\bf 09} (2020) 178,
  [\href{http://arxiv.org/abs/2007.14792}{{\tt arXiv:2007.14792}}].

\bibitem{Minkowski:1977sc}
P.~Minkowski, {\it {$\mu \to e\gamma$ at a Rate of One Out of $10^{9}$ Muon
  Decays?}},  Phys. Lett. B {\bf 67} (1977) 421--428.

\bibitem{Gell-Mann:1979vob}
M.~Gell-Mann, P.~Ramond, and R.~Slansky, {\it {Complex Spinors and Unified
  Theories}},  Conf. Proc. C {\bf 790927} (1979) 315--321,
  [\href{http://arxiv.org/abs/1306.4669}{{\tt arXiv:1306.4669}}].

\bibitem{Yanagida:1979as}
T.~Yanagida, {\it {Horizontal gauge symmetry and masses of neutrinos}},  C.P.C
  {\bf 7902131} (1979) 95--99.

\bibitem{Mohapatra:1979ia}
R.~N. Mohapatra and G.~Senjanovic, {\it {Neutrino Mass and Spontaneous Parity
  Nonconservation}},  Phys. Rev. Lett. {\bf 44} (1980) 912.

\bibitem{Langacker:1988ur}
P.~Langacker and D.~London, {\it {Mixing Between Ordinary and Exotic
  Fermions}},  Phys. Rev. D {\bf 38} (1988) 886.

\bibitem{Antusch:2006vwa}
S.~Antusch, C.~Biggio, E.~Fernandez-Martinez, M.~B. Gavela, and J.~Lopez-Pavon,
  {\it {Unitarity of the Leptonic Mixing Matrix}},  JHEP {\bf 10} (2006) 084,
  [\href{http://arxiv.org/abs/hep-ph/0607020}{{\tt hep-ph/0607020}}].

\bibitem{Antusch:2014woa}
S.~Antusch and O.~Fischer, {\it {Non-unitarity of the leptonic mixing matrix:
  Present bounds and future sensitivities}},  JHEP {\bf 10} (2014) 094,
  [\href{http://arxiv.org/abs/1407.6607}{{\tt arXiv:1407.6607}}].

\bibitem{Fernandez-Martinez:2016lgt}
E.~Fernandez-Martinez, J.~Hernandez-Garcia, and J.~Lopez-Pavon, {\it {Global
  constraints on heavy neutrino mixing}},  JHEP {\bf 08} (2016) 033,
  [\href{http://arxiv.org/abs/1605.08774}{{\tt arXiv:1605.08774}}].

\bibitem{UTfit:2022hsi}
{\bf UTfit} Collaboration, M.~Bona {\em et.~al.}, {\it {New UTfit Analysis of
  the Unitarity Triangle in the Cabibbo-Kobayashi-Maskawa scheme}},
  \href{http://arxiv.org/abs/2212.03894}{{\tt arXiv:2212.03894}}.

\bibitem{Muong-2:2006rrc}
{\bf Muon $g-2$} Collaboration, G.~W. Bennett {\em et.~al.}, {\it {Final Report
  of the Muon E821 Anomalous Magnetic Moment Measurement at BNL}},  Phys. Rev.
  D {\bf 73} (2006) 072003, [\href{http://arxiv.org/abs/hep-ex/0602035}{{\tt
  hep-ex/0602035}}].

\bibitem{Muong-2:2021ojo}
{\bf Muon $g-2$} Collaboration, B.~Abi {\em et.~al.}, {\it {Measurement of the
  Positive Muon Anomalous Magnetic Moment to 0.46 ppm}},  Phys. Rev. Lett. {\bf
  126} (2021), no.~14 141801, [\href{http://arxiv.org/abs/2104.03281}{{\tt
  arXiv:2104.03281}}].

\bibitem{Aoyama:2020ynm}
T.~Aoyama {\em et.~al.}, {\it {The anomalous magnetic moment of the muon in the
  Standard Model}},  Phys. Rept. {\bf 887} (2020) 1--166,
  [\href{http://arxiv.org/abs/2006.04822}{{\tt arXiv:2006.04822}}].

\bibitem{Borsanyi:2020mff}
S.~Borsanyi {\em et.~al.}, {\it {Leading hadronic contribution to the muon
  magnetic moment from lattice QCD}},  Nature {\bf 593} (2021), no.~7857
  51--55, [\href{http://arxiv.org/abs/2002.12347}{{\tt arXiv:2002.12347}}].

\bibitem{Ce:2022kxy}
M.~Ce {\em et.~al.}, {\it {Window observable for the hadronic vacuum
  polarization contribution to the muon $g-2$ from lattice QCD}},
  \href{http://arxiv.org/abs/2206.06582}{{\tt arXiv:2206.06582}}.

\bibitem{Alexandrou:2022amy}
C.~Alexandrou {\em et.~al.}, {\it {Lattice Calculation of the Short and
  Intermediate Time-Distance Hadronic Vacuum Polarization Contributions to the
  Muon Magnetic Moment Using Twisted-Mass Fermions}},
  \href{http://arxiv.org/abs/2206.15084}{{\tt arXiv:2206.15084}}.

\bibitem{Athron:2022qpo}
P.~Athron, A.~Fowlie, C.-T. Lu, L.~Wu, Y.~Wu, and B.~Zhu, {\it {The $W$ boson
  Mass and Muon $g-2$: Hadronic Uncertainties or New Physics?}},
  \href{http://arxiv.org/abs/2204.03996}{{\tt arXiv:2204.03996}}.

\bibitem{Crivellin:2020zul}
A.~Crivellin, M.~Hoferichter, C.~A. Manzari, and M.~Montull, {\it {Hadronic
  Vacuum Polarization: $(g-2)_\mu$ versus Global Electroweak Fits}},  Phys.
  Rev. Lett. {\bf 125} (2020), no.~9 091801,
  [\href{http://arxiv.org/abs/2003.04886}{{\tt arXiv:2003.04886}}].

\bibitem{Keshavarzi:2020bfy}
A.~Keshavarzi, W.~J. Marciano, M.~Passera, and A.~Sirlin, {\it {Muon $g-2$ and
  $\Delta \alpha$ connection}},  Phys. Rev. D {\bf 102} (2020), no.~3 033002,
  [\href{http://arxiv.org/abs/2006.12666}{{\tt arXiv:2006.12666}}].

\bibitem{Malaescu:2020zuc}
B.~Malaescu and M.~Schott, {\it {Impact of correlations between $a_{\mu}$ and
  $\alpha_\text{QED}$ on the EW fit}},  Eur. Phys. J. C {\bf 81} (2021), no.~1
  46, [\href{http://arxiv.org/abs/2008.08107}{{\tt arXiv:2008.08107}}].

\bibitem{Colangelo:2020lcg}
G.~Colangelo, M.~Hoferichter, and P.~Stoffer, {\it {Constraints on the two-pion
  contribution to hadronic vacuum polarization}},  Phys. Lett. B {\bf 814}
  (2021) 136073, [\href{http://arxiv.org/abs/2010.07943}{{\tt
  arXiv:2010.07943}}].

\bibitem{Kannike:2011ng}
K.~Kannike, M.~Raidal, D.~M. Straub, and A.~Strumia, {\it {Anthropic solution
  to the magnetic muon anomaly: the charged see-saw}},  JHEP {\bf 02} (2012)
  106, [\href{http://arxiv.org/abs/1111.2551}{{\tt arXiv:1111.2551}}].
  [Erratum: JHEP 10, 136 (2012)].

\bibitem{Dermisek:2013gta}
R.~Dermisek and A.~Raval, {\it {Explanation of the Muon $g-2$ Anomaly with
  Vectorlike Leptons and its Implications for Higgs Decays}},  Phys. Rev. D
  {\bf 88} (2013) 013017, [\href{http://arxiv.org/abs/1305.3522}{{\tt
  arXiv:1305.3522}}].

\bibitem{Arcadi:2021cwg}
G.~Arcadi, L.~Calibbi, M.~Fedele, and F.~Mescia, {\it {Muon $g-2$ and
  $B$-anomalies from Dark Matter}},  Phys. Rev. Lett. {\bf 127} (2021), no.~6
  061802, [\href{http://arxiv.org/abs/2104.03228}{{\tt arXiv:2104.03228}}].

\bibitem{Lu:2021vcp}
C.-T. Lu, R.~Ramos, and Y.-L.~S. Tsai, {\it {Shedding light on dark matter with
  recent muon $(g-2)$ and Higgs exotic decay measurements}},  JHEP {\bf 08}
  (2021) 073, [\href{http://arxiv.org/abs/2104.04503}{{\tt arXiv:2104.04503}}].

\bibitem{Guedes:2022cfy}
G.~Guedes and P.~Olgoso, {\it {A bridge to new physics: proposing new -- and
  reviving old -- explanations of $a_\mu$}},  JHEP {\bf 09} (2022) 181,
  [\href{http://arxiv.org/abs/2205.04480}{{\tt arXiv:2205.04480}}].

\bibitem{Arkani-Hamed:2021xlp}
N.~Arkani-Hamed and K.~Harigaya, {\it {Naturalness and the muon magnetic
  moment}},  JHEP {\bf 09} (2021) 025,
  [\href{http://arxiv.org/abs/2106.01373}{{\tt arXiv:2106.01373}}].

\bibitem{Craig:2021ksw}
N.~Craig, I.~G. Garcia, A.~Vainshtein, and Z.~Zhang, {\it {Magic zeroes and
  hidden symmetries}},  JHEP {\bf 05} (2022) 079,
  [\href{http://arxiv.org/abs/2112.05770}{{\tt arXiv:2112.05770}}].

\bibitem{DelleRose:2022ygn}
L.~Delle~Rose, B.~von Harling, and A.~Pomarol, {\it {Wilson coefficients and
  natural zeros from the on-shell viewpoint}},  JHEP {\bf 05} (2022) 120,
  [\href{http://arxiv.org/abs/2201.10572}{{\tt arXiv:2201.10572}}].

\bibitem{ATLAS:2019nkf}
{\bf ATLAS} Collaboration, G.~Aad {\em et.~al.}, {\it {Combined measurements of
  Higgs boson production and decay using up to $80$ fb$^{-1}$ of proton-proton
  collision data at $\sqrt{s}=$ 13 TeV collected with the ATLAS experiment}},
  Phys. Rev. D {\bf 101} (2020), no.~1 012002,
  [\href{http://arxiv.org/abs/1909.02845}{{\tt arXiv:1909.02845}}].

\bibitem{CMS:2018uag}
{\bf CMS} Collaboration, A.~M. Sirunyan {\em et.~al.}, {\it {Combined
  measurements of Higgs boson couplings in proton-proton collisions at
  $\sqrt{s}=13$ TeV}},  Eur. Phys. J. C {\bf 79} (2019), no.~5 421,
  [\href{http://arxiv.org/abs/1809.10733}{{\tt arXiv:1809.10733}}].

\bibitem{ATLAS:2020fzp}
{\bf ATLAS} Collaboration, G.~Aad {\em et.~al.}, {\it {A search for the dimuon
  decay of the Standard Model Higgs boson with the ATLAS detector}},  Phys.
  Lett. B {\bf 812} (2021) 135980, [\href{http://arxiv.org/abs/2007.07830}{{\tt
  arXiv:2007.07830}}].

\bibitem{CMS:2020xwi}
{\bf CMS} Collaboration, A.~M. Sirunyan {\em et.~al.}, {\it {Evidence for Higgs
  boson decay to a pair of muons}},  JHEP {\bf 01} (2021) 148,
  [\href{http://arxiv.org/abs/2009.04363}{{\tt arXiv:2009.04363}}].

\bibitem{Alonso-Gonzalez:2021tpo}
J.~Alonso-Gonzalez, A.~de~Giorgi, L.~Merlo, and S.~Pokorski, {\it {Searching
  for BSM physics in Yukawa couplings and flavour symmetries}},  JHEP {\bf 05}
  (2022) 041, [\href{http://arxiv.org/abs/2109.07490}{{\tt arXiv:2109.07490}}].

\bibitem{Bahl:2022yrs}
H.~Bahl, E.~Fuchs, S.~Heinemeyer, J.~Katzy, M.~Menen, K.~Peters, M.~Saimpert,
  and G.~Weiglein, {\it {Constraining the CP structure of Higgs-fermion
  couplings with a global LHC fit, the electron EDM and baryogenesis}},  Eur.
  Phys. J. C {\bf 82} (2022), no.~7 604,
  [\href{http://arxiv.org/abs/2202.11753}{{\tt arXiv:2202.11753}}].

\bibitem{Brod:2022bww}
J.~Brod, J.~M. Cornell, D.~Skodras, and E.~Stamou, {\it {Global constraints on
  Yukawa operators in the standard model effective theory}},  JHEP {\bf 08}
  (2022) 294, [\href{http://arxiv.org/abs/2203.03736}{{\tt arXiv:2203.03736}}].

\bibitem{Breso-Pla:2021qoe}
V.~Breso-Pla, A.~Falkowski, and M.~Gonzalez-Alonso, {\it {$A_{FB}$ in the
  SMEFT: precision Z physics at the LHC}},  JHEP {\bf 08} (2021) 021,
  [\href{http://arxiv.org/abs/2103.12074}{{\tt arXiv:2103.12074}}].

\bibitem{Workman:2022ynf}
{\bf Particle Data Group} Collaboration, R.~L. Workman, {\it {Review of
  Particle Physics}},  PTEP {\bf 2022} (2022) 083C01.

\bibitem{Janot:2019oyi}
P.~Janot and S.~Jadach, {\it {Improved Bhabha cross section at LEP and the
  number of light neutrino species}},  Phys. Lett. B {\bf 803} (2020) 135319,
  [\href{http://arxiv.org/abs/1912.02067}{{\tt arXiv:1912.02067}}].

\bibitem{Bryman:2021teu}
D.~Bryman, V.~Cirigliano, A.~Crivellin, and G.~Inguglia, {\it {Testing Lepton
  Flavor Universality with Pion, Kaon, Tau, and Beta Decays}},
  \href{http://arxiv.org/abs/2111.05338}{{\tt arXiv:2111.05338}}.

\bibitem{Abdullahi:2022jlv}
A.~M. Abdullahi {\em et.~al.}, {\it {The Present and Future Status of Heavy
  Neutral Leptons}},  in {\em {2022 Snowmass Summer Study}}, 3, 2022.
\newblock \href{http://arxiv.org/abs/2203.08039}{{\tt arXiv:2203.08039}}.

\bibitem{Freitas:2014pua}
A.~Freitas, J.~Lykken, S.~Kell, and S.~Westhoff, {\it {Testing the Muon $g-2$
  Anomaly at the LHC}},  JHEP {\bf 05} (2014) 145,
  [\href{http://arxiv.org/abs/1402.7065}{{\tt arXiv:1402.7065}}]. [Erratum:
  JHEP 09, 155 (2014)].

\bibitem{Huang:2020hdv}
G.-y. Huang and S.~Zhou, {\it {Precise Values of Running Quark and Lepton
  Masses in the Standard Model}},  Phys. Rev. D {\bf 103} (2021), no.~1 016010,
  [\href{http://arxiv.org/abs/2009.04851}{{\tt arXiv:2009.04851}}].

\bibitem{Arason:1991ic}
H.~Arason, D.~J. Castano, B.~Keszthelyi, S.~Mikaelian, E.~J. Piard, P.~Ramond,
  and B.~D. Wright, {\it {Renormalization group study of the standard model and
  its extensions. 1. The Standard model}},  Phys. Rev. D {\bf 46} (1992)
  3945--3965.

\bibitem{Gavela:2009cd}
M.~B. Gavela, T.~Hambye, D.~Hernandez, and P.~Hernandez, {\it {Minimal Flavour
  Seesaw Models}},  JHEP {\bf 09} (2009) 038,
  [\href{http://arxiv.org/abs/0906.1461}{{\tt arXiv:0906.1461}}].

\end{thebibliography}\endgroup
\bibliographystyle{BiblioStyle}

\end{document}